\documentclass[ALICE,manyauthors]{cernphprep}
\usepackage[comma,square,numbers,sort&compress]{natbib}

\usepackage{lineno}
\usepackage{xspace}
\usepackage{hyperref}
\usepackage{siunitx}
\usepackage{upgreek}

\usepackage[T1]{fontenc}
\usepackage{orcidlink}

\begin{document}

\newcommand{\pel}{\ensuremath{\rm e}\xspace}
\newcommand{\pmu}{\ensuremath{\mu}\xspace}
\newcommand{\pde}{\ensuremath{\rm d}\xspace}
\newcommand{\dbar}{\ensuremath{\rm \bar{d}}\xspace}
\newcommand{\ppip}{\ensuremath{\pi^{+}}\xspace}
\newcommand{\ppim}{\ensuremath{\pi^{-}}\xspace}
\newcommand{\ppi}{\ensuremath{\pi}\xspace}
\newcommand{\ppipm}{\ensuremath{\pi^{\pm}}\xspace}
\newcommand{\ppis}{\ensuremath{\ppip + \ppim}\xspace}
\newcommand{\pkap}{\ensuremath{{\rm K}^{+}}\xspace}
\newcommand{\pkam}{\ensuremath{{\rm K}^{-}}\xspace}
\newcommand{\pka}{\ensuremath{{\rm K}}\xspace}
\newcommand{\pkapm}{\ensuremath{{\rm K}^{\pm}}\xspace}
\newcommand{\pkas}{\ensuremath{\pkap + \pkam}\xspace}
\newcommand{\pkastar}{\ensuremath{{\rm K}^{*}}\xspace}
\newcommand{\pkastarm}{\ensuremath{\overline{\pkastar}}\xspace}
\newcommand{\pkazero}{\ensuremath{{\rm K}^{0}_{\rm S}}\xspace}
\newcommand{\pprp}{\ensuremath{{\rm p}}\xspace}
\newcommand{\pprm}{\ensuremath{\overline{\rm{p}}}\xspace}
\newcommand{\ppr}{\ensuremath{{\rm p}}\xspace}
\newcommand{\pprpm}{\ensuremath{(\pprm) \pprp }\xspace}
\newcommand{\pprs}{\ensuremath{\pprp + \pprm}\xspace}
\newcommand{\pdeu}{\ensuremath{{\rm d}}\xspace}
\newcommand{\pphi}{\ensuremath{\phi}\xspace}
\newcommand{\plam}{\ensuremath{\Lambda}\xspace}
\newcommand{\psigp}{\ensuremath{\Sigma^{+}}\xspace}
\newcommand{\rppi}{\ensuremath{\ppr/\ppi}\xspace}
\newcommand{\rphipi}{\ensuremath{\pphi/\ppi}\xspace}

\newcommand{\Mmunu}{\ensuremath{{\rm M}(\mu \nu_{\mu})}\xspace}
\newcommand{\ndfITS}{\ensuremath{N^{\rm hits}_{\rm ITS}}\xspace}
\newcommand{\chitwo}{\ensuremath{\chi^{2}}\xspace}
\newcommand{\chitwoITS}{\ensuremath{\chi^{2}_{\rm ITS}}\xspace}
\newcommand{\chitwoTPC}{\ensuremath{\chi^{2}_{\rm TPC}}\xspace}
\newcommand{\chitwoNDF}{\ensuremath{\chi^{2}/{\rm NDF}}\xspace}
\newcommand{\dca}{\ensuremath{{\rm DCA}}\xspace}
\newcommand{\dcaXY}{\ensuremath{\dca_{\it{xy}}}\xspace}
\newcommand{\sigmadcaXY}{\ensuremath{\sigma_{\dcaXY}}\xspace}
\newcommand{\dcaZ}{\ensuremath{\dca_{\it{z}}}\xspace}
\newcommand{\mom}{\ensuremath{p}\xspace}
\newcommand{\momTRUE}{\ensuremath{\mom^{\rm TRUE}}\xspace}
\newcommand{\momMEAS}{\ensuremath{\mom^{\rm MEAS}}\xspace}
\newcommand{\pt}{\ensuremath{p_{\rm{T}}}\xspace}
\newcommand{\ptTRUE}{\ensuremath{\pt^{\rm TRUE}}\xspace}
\newcommand{\ptMEAS}{\ensuremath{\pt^{\rm MEAS}}\xspace}
\newcommand{\meanpt}{\ensuremath{\langle \pt \rangle}\xspace}
\newcommand{\mt}{\ensuremath{m_{\rm{T}}}\xspace}
\newcommand{\qt}{\ensuremath{q_{\rm{T}}}\xspace}
\newcommand{\cmc}{\ensuremath{\text{cm}/c}\xspace}
\newcommand{\texpi}{\ensuremath{t_{\rm exp}(i)}\xspace}
\newcommand{\texp}{\ensuremath{t_{\rm exp}}\xspace}
\newcommand{\evtime}{\ensuremath{t_{\rm ev}}\xspace}
\newcommand{\ttof}{\ensuremath{t_{\rm TOF}}\xspace}
\newcommand{\Nsigma}{\ensuremath{{\rm N}_{\sigma}}\xspace}
\newcommand{\nsigma}{\ensuremath{\Nsigma}\xspace}
\newcommand{\Nsigmai}{\ensuremath{{\rm N}_{\sigma, i}}\xspace}
\newcommand{\NsigmaTPC}{\ensuremath{{\rm N}_{\sigma}^{\rm TPC}}\xspace}
\newcommand{\NsigmaTOF}{\ensuremath{{\rm N}_{\sigma}^{\rm TOF}}\xspace}
\newcommand{\tofmeas}{\ensuremath{\text{time-of-flight}}\xspace}
\newcommand{\AAaa}{\ensuremath{{\rm{AA}}}\xspace}
\newcommand{\raa}{\ensuremath{R_{\AAaa}}\xspace}
\newcommand{\taa}{\ensuremath{T_{\AAaa}}\xspace}
\newcommand{\ncoll}{\ensuremath{N_{\rm{coll}}}\xspace}
\newcommand{\dedx}{\ensuremath{{\rm d}E/{\rm d}x}\xspace}
\newcommand{\meandedx}{\ensuremath{\langle\dedx\rangle}\xspace}
\newcommand{\dndyV}{\ensuremath{{\rm d}N/{\rm d}y}\xspace}
\newcommand{\Nch}{\ensuremath{{\it N}_{\rm{ch}}}\xspace}
\newcommand{\dNchdeta}{\ensuremath{\rm{d}\Nch/\rm{d}\eta}\xspace}
\newcommand{\avdNchdeta}{\ensuremath{\langle\dNchdeta\rangle}\xspace}
\newcommand{\dndydpt}{\ensuremath{{\rm d}^{2}N/{\rm d}y{\rm d}\pt}\xspace}
\newcommand{\Tch}{\ensuremath{T_{\rm{ch}}}\xspace}
\newcommand{\Tc}{\ensuremath{T_{\rm{c}}}\xspace}
\newcommand{\Tkin}{\ensuremath{T_{\rm{kin}}}\xspace}
\newcommand{\Bt}{\ensuremath{\beta_{\rm{T}}}\xspace}
\newcommand{\avBt}{\ensuremath{\langle\Bt\rangle}\xspace}
\newcommand{\EcrossB}{\ensuremath{E\times B}\xspace}
\newcommand{\centint}[2]{\ensuremath{#1-#2\%}\xspace}
\newcommand{\momint}[2]{\ensuremath{#1-#2~\gevc}\xspace}
\newcommand{\minv}{\ensuremath{M_{\rm{KK}}}\xspace}
\newcommand{\twopi}{\ensuremath{2\pi}\xspace}
\newcommand{\fun}[2]{\ensuremath{#1\left(#2\right)}}

\newcommand{\xexe}{\ensuremath{\text{Xe--Xe}}\xspace}
\newcommand{\pbpb}{\ensuremath{\text{Pb--Pb}}\xspace}
\newcommand{\ppb}{\ensuremath{\text{p--Pb}}\xspace}
\newcommand{\pp}{\ensuremath{\text{pp}}\xspace}
\newcommand{\gevcV}{\ensuremath{{\rm GeV}/c}\xspace}
\newcommand{\mevcV}{\ensuremath{{\rm MeV}/c}\xspace}
\newcommand{\tevcV}{\ensuremath{{\rm TeV}/c}\xspace}
\newcommand{\gevcsq}{\ensuremath{{\gevcV}^{2}}\xspace}

\newcommand{\sF}{\ensuremath{\s~=~5.02}\xspace}
\newcommand{\snn}{\ensuremath{\sqrt{s_{\rm{NN}}}}\xspace}
\newcommand{\snnT}{\ensuremath{\snn~=~2.76}\xspace}
\newcommand{\snnF}{\ensuremath{\snn~=~5.02}\xspace}
\newcommand{\snnXeXe}{\ensuremath{\snn~=~5.44}\xspace~TeV~}

\newcommand{\Refs}[1]{Refs.~\cite{#1}\xspace}
\newcommand{\Tab}[1]{Tab.~\ref{#1}\xspace}
\newcommand{\Eq}[1]{Eq.~\ref{#1}\xspace}
\newcommand{\Figs}[1]{Figs.~\ref{#1}\xspace}
\newcommand{\Fig}[1]{Fig.~\ref{#1}\xspace}
\newcommand{\Sec}[1]{Section~\ref{#1}\xspace}
\newcommand{\Ustat}{Stat. Uncert.\xspace}
\newcommand{\Usyst}{Syst. Uncert.\xspace}
\newcommand{\BR}{B.R.\xspace}

\newcommand{\GEANTT}{\ensuremath{\text{GEANT3}}\xspace}
\newcommand{\GEANTF}{\ensuremath{\text{GEANT4}}\xspace}
\newcommand{\FLUKA}{\ensuremath{\text{FLUKA}}\xspace}
\newcommand{\HIJING}{\ensuremath{\text{HIJING}}\xspace}

\newcommand{\Raa}{\ensuremath{\rm R_{AA}}\xspace}
\newcommand{\Rpa}{\ensuremath{\rm R_{pA}}\xspace}
\newcommand{\Rppb}{\ensuremath{R_{\rm pPb}}\xspace}

\newcommand{\cmsquared}{\ensuremath{{\rm cm}^{2}}\xspace}
\newcommand{\cme}{\ensuremath{\sqrt{s}}\xspace}
\newcommand{\gev}{\ensuremath{{\rm GeV}/c}\xspace}
\newcommand{\auau}{\ensuremath{\rm Au\!-\!Au}\xspace}
\newcommand{\ptopi}{\ensuremath{{\rm p } / \pi}\xspace}
\newcommand{\ktopi}{\ensuremath{({\rm K}^{+}+{\rm K}^{-}) / (\pi^{+}+\pi^{-})}\xspace}
\newcommand{\twotwo}{\ensuremath{2\rightarrow 2}\xspace}
\newcommand{\ltok}{\ensuremath{({\rm \Lambda}^{0}+\bar {\rm \Lambda}^{0})/(\rm 2 K^{0}_{s} )} \xspace}

\newcommand{\snnt}[1]{\ensuremath{\snn~=~#1~\text{\,TeV}}\xspace}
\newcommand{\snnnotext}[1]{\ensuremath{\snn~=~#1}\xspace}
\newcommand{\sppt}[1]{\ensuremath{\sqrt{s} = #1 \text{\,TeV}}\xspace}
\newcommand{\sppg}[1]{\ensuremath{\sqrt{s} = #1 \text{\,GeV}}\xspace}
\newcommand{\gevc}[1]{\ensuremath{#1\ \text{\,\gevcV}}\xspace}
\newcommand{\mevc}[1]{\ensuremath{#1\ \text{\,\mevcV}}\xspace}
\newcommand{\mevcsq}[1]{\ensuremath{#1\ \text{\,\mevcV}^{2}}\xspace}
\newcommand{\dndeta}[1]{\ensuremath{\frac{\text{d}^2N_{#1}}{\text{d}\pt \text{d}\eta}}\xspace}
\newcommand{\dndy}[1]{\ensuremath{\frac{d^2N_{#1}}{d\pt dy}}\xspace}
\newcommand{\eff}[1]{\ensuremath{\epsilon_{#1}}\xspace}
\newcommand{\bareyield}{\ensuremath{Y}\xspace}
\newcommand{\yield}[1]{\ensuremath{Y_{#1}}\xspace}
\newcommand{\ch}{\ensuremath{\text{ch}}\xspace}
\newcommand{\etalab}{\ensuremath{\eta_{lab}}\xspace}
\newcommand{\etaint}[1]{\ensuremath{|\eta| < #1}\xspace}
\newcommand{\yint}[1]{\ensuremath{|y| < #1}\xspace}

\newcommand{\VZ}{\ensuremath{V^{0}}\xspace}
\newcommand{\VZs}{\ensuremath{V^{0}\text{s}}\xspace}
\newcommand{\RAA}{\ensuremath{R_{\text{AA}}}\xspace}
\newcommand{\MRAA}{\ensuremath{\mathbf{R_{\text{\textbf{AA}}}}}\xspace}
\newcommand{\dpi}{\ensuremath{\Delta_{\pi}}\xspace}
\newcommand{\dkaon}{\ensuremath{\Delta_{K}}\xspace}
\newcommand{\dproton}{\ensuremath{\Delta_{p}}\xspace}
\newcommand{\mdpi}{\ensuremath{\mathbf{\Delta_{\pi}}}\xspace}
\newcommand{\mdkaon}{\ensuremath{\mathbf{\Delta_{K}}}\xspace}
\newcommand{\mdproton}{\ensuremath{\mathbf{\Delta_{p}}}\xspace}
\newcommand{\rpi}{\ensuremath{R_{\pi}}\xspace}
\newcommand{\mathdedx}{\ensuremath{\mathbf{\text{d}E/\text{d}x}}\xspace}
\newcommand{\mdedx}{\ensuremath{\left <\text{d}E/\text{d}x \right>}\xspace}
\newcommand{\mathmdedx}{\ensuremath{\mathbf{\left <\text{d}E/\text{d}x \right>}}\xspace}
\newcommand{\mdedxpi}{\ensuremath{\left <\text{d}E/\text{d}x \right>_{\pi}}\xspace}
\newcommand{\meanp}{\ensuremath{\langle p \rangle}\xspace}
\newcommand{\sdedx}{\ensuremath{\sigma_{\text{d}E/\text{d}x}}\xspace}
\newcommand{\relres}{\ensuremath{\sigma/\left <\text{d}E/\text{d}x \right>}\xspace}
\newcommand{\res}{\ensuremath{\sigma_{\text{d}E/\text{d}x}}\xspace}
\newcommand{\ncl}{\ensuremath{\text{Ncl}}\xspace}
\newcommand{\mncl}{\ensuremath{\langle \text{Ncl} \rangle}\xspace}

\newcommand{\chpi}{\ensuremath{\pi^{+}+\pi^{-}}\xspace}
\newcommand{\chk}{\ensuremath{{\rm K}^{+}+{\rm K}^{-}}\xspace}
\newcommand{\chp}{\ensuremath{{\rm p}+{\rm \bar{p}}}\xspace}

\newcommand{\bg}{\ensuremath{\beta\gamma}\xspace}

\newcommand{\timemeasure}{\ensuremath{{t}\xspace}}
\newcommand{\timemeasureTOF}{\ensuremath{{{\timemeasure}_{\rm TOF}}\xspace}}
\newcommand{\Otwo}{\ensuremath{{\rm O}^{2}}\xspace}
\newcommand{\LA}{\ensuremath{\Lambda}\xspace}
\newcommand{\AL}{\ensuremath{\bar{\Lambda}}\xspace}
\newcommand{\KOs}{\ensuremath{\rm K^0_S}\xspace}
\newcommand{\dMassG}{\ensuremath{\Delta m_\gamma}\xspace}
\newcommand{\dMassL}{\ensuremath{\Delta m_\Lambda}\xspace}
\newcommand{\dMassAL}{\ensuremath{\Delta m_{\bar{\Lambda}}}\xspace}
\newcommand{\dMassKO}{\ensuremath{\Delta m_{K^0_s}}\xspace}
\newcommand{\TFF}{\ensuremath{\texttt{TFractionFitter}}\xspace}
\newcommand{\RooFit}{\ensuremath{\texttt{RooFit}}\xspace}
\newcommand{\deltaOne}{\ensuremath{{(\deltaPi)_{ \rm 1}}}\xspace}
\newcommand{\deltaTwo}{\ensuremath{{(\deltaPi)_{ \rm 2}}}\xspace}
\newcommand{\sigmatoftrk}{\ensuremath{{\rm \sigma_{TOF} \oplus \sigma_{Trk}}}\xspace}
\newcommand{\doubledelta}{\ensuremath{{\rm \Delta \Delta} \timemeasureTOF}\xspace}
\newcommand{\sigmadoubledelta}{\ensuremath{{\sigma_{\doubledelta}}}\xspace}
\newcommand{\sigmadoubledeltaref}{\ensuremath{{\sigma_{\doubledelta}^{\rm Reference}}}\xspace}
\newcommand{\doubledeltaLong}{\ensuremath{\deltaTwo - \deltaOne}\xspace}
\newcommand{\texpPi}{\ensuremath{\texp(\uppi)}\xspace}
\newcommand{\texpKa}{\ensuremath{\texp(\rm K)}\xspace}
\newcommand{\texpPr}{\ensuremath{\texp(\rm p)}\xspace}
\newcommand{\deltaPi}{\ensuremath{\timemeasureTOF - \texpPi}\xspace}
\newcommand{\deltaKa}{\ensuremath{\timemeasureTOF - \texpKa}\xspace}
\newcommand{\deltaPr}{\ensuremath{\timemeasureTOF - \texpPr}\xspace}
\newcommand{\evTime}{\ensuremath{\timemeasure_{\rm ev}}\xspace}
\newcommand{\evTimeTOF}{\ensuremath{\evTime^{\rm TOF}}\xspace}
\newcommand{\evTimeTZAC}{\ensuremath{\evTime^{\rm FT0}}\xspace}
\newcommand{\evTimeTZA}{\ensuremath{\evTime^{\rm FT0A}}\xspace}
\newcommand{\evTimeTZC}{\ensuremath{\evTime^{\rm FT0C}}\xspace}
\newcommand{\sigmaTOF}{\ensuremath{ \sigma_{\rm TOF} }\xspace}
\newcommand{\sigmaTOFpid}{\ensuremath{ \sigmaTOF^{\rm PID} }\xspace}
\newcommand{\sigmaTrk}{\ensuremath{ \sigma_{\rm Tracking} }\xspace}
\newcommand{\sigmaRef}{\ensuremath{ \sigma_{\rm Reference} }\xspace}

\newcommand{\deltaGeneral}{\ensuremath{\timemeasureTOF - \texp - \evTime}\xspace}
\newcommand{\deltaGeneralTZAC}{\ensuremath{\timemeasureTOF - \texp - \evTimeTZAC}\xspace}
\newcommand{\deltaPiTOF}{\ensuremath{\timemeasureTOF - \texpPi - \evTimeTOF}\xspace}
\newcommand{\deltaKaTOF}{\ensuremath{\timemeasureTOF - \texpKa - \evTimeTOF}\xspace}
\newcommand{\deltaPrTOF}{\ensuremath{\timemeasureTOF - \texpPr - \evTimeTOF}\xspace}
\newcommand{\deltaPiTZAC}{\ensuremath{\timemeasureTOF - \texpPi - \evTimeTZAC}\xspace}
\newcommand{\deltaKaTZAC}{\ensuremath{\timemeasureTOF - \texpKa - \evTimeTZAC}\xspace}
\newcommand{\deltaPrTZAC}{\ensuremath{\timemeasureTOF - \texpPr - \evTimeTZAC}\xspace}
\newcommand{\tofEvMult}{\ensuremath{\rm TOF\ ev. mult.}\xspace}

\newcommand{\relval}{\texttt{rel\_val}\xspace}
\newcommand{\apassthree}{\texttt{apass3}\xspace}
\newcommand{\RunTwo}{\ensuremath{\rm Run~2}\xspace}
\newcommand{\RunThree}{\ensuremath{\rm Run~3}\xspace}
\newcommand{\trkOne}{\ensuremath{\rm track_{1}}\xspace}
\newcommand{\trkTwo}{\ensuremath{\rm track_{2}}\xspace}


\newcommand{\DoubleDeltaRefSigmaValue}{\ensuremath{80.6}\xspace}
\newcommand{\DoubleDeltaRefSigmaError}{\ensuremath{0.4}\xspace}
\newcommand{\DoubleDeltaResultValue}{\ensuremath{74.7}\xspace}
\newcommand{\DoubleDeltaResult}{\ensuremath{\DoubleDeltaResultValue \pm 2.3\ {\rm ps}}\xspace}

\newcommand{\StdResultValue}{\ensuremath{72.8}\xspace}
\newcommand{\StdResult}{\ensuremath{\StdResultValue \pm 0.5\ {\rm ps}}\xspace}
\newcommand{\StdResultFullEta}{\ensuremath{77.9 \pm 0.8\ {\rm ps}}\xspace}
  \newcommand{\FTZResoValue}{\ensuremath{16\ {\rm ps}}\xspace}

\begin{titlepage}
\PHyear{2025}       
\PHnumber{255}      
\PHdate{04 November}  

\title{Time resolution of the ALICE Time-Of-Flight detector with the first Run~3 pp collisions at ${\bf \sqrt{\textit{s}} = 13.6}$ TeV}
\ShortTitle{Time resolution of the ALICE Time-Of-Flight detector in Run~3}   

\Collaboration{ALICE Collaboration\thanks{See Appendix~\ref{app:collab} for the list of collaboration members}}
\ShortAuthor{ALICE Collaboration} 

\begin{abstract}
Particle identification (PID) is a fundamental aspect of the ALICE detector system, central to its heavy-ion and proton-proton physics programs.
Among the different PID strategies, ALICE uses the Time-Of-Flight (TOF) detector to identify particles at intermediate momenta ($0.5 < \pt <\gevc{4}$).
The ALICE TOF detector performed successfully during the first ten years of LHC operations.
During the Long Shutdown 2, many ALICE sub-detectors, including TOF, were upgraded to fully leverage the targeted 50 kHz interaction rate of Pb--Pb collisions, which required the implementation of a continuous readout scheme.
The TOF detector electronics were upgraded and refurbished, while processing algorithms for data quality control, reconstruction, calibration, and analysis were rewritten.
This paper presents the upgraded TOF detector operation and calibration procedures and its performance in terms of timing resolution, a key factor for particle separation in ALICE analyses.
Using 2022 pp collision data at $\sqrt{s} = 13.6\ {\rm TeV}$ from Run~3, the time resolution of the detector was estimated with two independent methods, both yielding consistent results, better than 80~ps.
Despite the excellent performance already achieved, further improvements are expected after additional detector commissioning and refined calibration procedures, thus enhancing the ALICE PID capabilities for Run~3 and beyond.

\end{abstract}
\vspace{1.5cm}
\begin{center}
\emph{The successful deployment and performance of the ALICE Time-Of-Flight detector stand \\as a testament to the foundational contributions of our late colleague, Prof. Andrea Alici.\\This publication is dedicated to his memory.}
\end{center}

\end{titlepage}

\setcounter{page}{2} 


\section{Introduction}\label{sec:pap:introduction}

The core physics program of the ALICE experiment at the LHC is the study of the properties of the strongly interacting, dense, and hot matter created in high-energy heavy-ion collisions.
Most of the physics analyses rely on the capability to perform a precise particle identification (PID).
The PID strategy applied by the ALICE collaboration is based on the adoption of complementary techniques to ensure a wide coverage in momentum, from $\sim 100 \mevc$ to $\sim 20 \gevc$~\cite{ALICE:2019hno}.
The time-of-flight measurement with the ALICE TOF detector system provides the separation of charged particles in the intermediate transverse momentum range (from 0.5 to \gevc{4}).
This is achieved by ensuring a precise determination of the event collision time, the track length, and the particle's momentum, together with a precise measurement of the arrival time of the particles at the TOF detector.
At the end of LHC Run 2 (2018), the TOF detector successfully reached more than 10 years of operation.
During the Long Shutdown 2 (LS2, 2019-2021), the ALICE Collaboration upgraded its detector to cope with the higher interaction rate expected to be delivered by the LHC by adopting a continuous readout scheme.
An upgrade of the TOF electronics was performed to make the data acquisition (DAQ) system compliant with the new paradigm of ALICE data taking.
A full refurbishment of several of its components was also carried out, together with an upgrade of the offline data processing algorithms (data quality control, reconstruction, calibration, and analysis).
This article reports on the performance of the upgraded ALICE TOF detector obtained with Run~3 data, focusing on the achieved timing resolution.

\subsection{The ALICE detector in Run~3}
\begin{figure}
  \centering
  \includegraphics[width=0.9\textwidth]{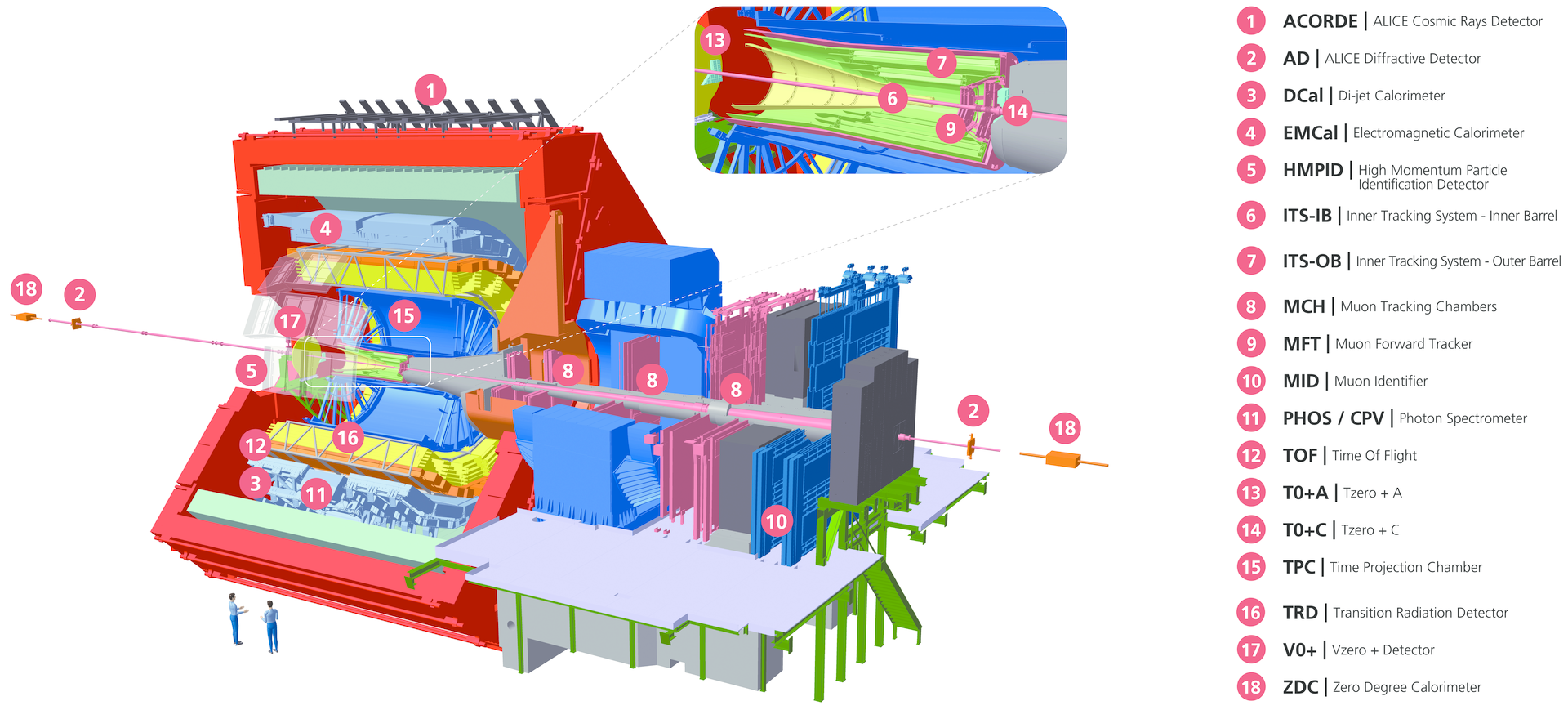}
  \caption{
    Schematic of the ALICE detector in Run~3.
  }
  \label{fig:AliceExp}
\end{figure}

Based on the experience developed in Run~1 and Run~2, the ALICE detector underwent a series of upgrades during the LS2 to enable future advancements focused on precision measurements of statistics-limited rare probes in both large (Pb--Pb) and small collision systems (pp)~\cite{Citron:2018lsq}.
In \Fig{fig:AliceExp},  the upgraded ALICE detector used in Run~3 is depicted: the apparatus couples an improved tracking performance together with capabilities to sustain data taking in high interaction rate conditions up to 1~MHz in pp collisions and 50~kHz in Pb--Pb collisions.
In the reference coordinate system of ALICE, the $z$-axis, defining the direction along which the pseudorapidity $\eta$ is measured, is given by the beam direction, and the azimuthal angle $\varphi$ is defined in the orthogonal plane.
In the following, a brief description of the detectors relevant for the measurement reported in this article will be given.
The Inner Tracking System (ITS)~\cite{ALICE:2013nwm} is composed of seven cylindrical layers of silicon pixel detectors located at radial distances between 2.3 and 39.3~cm from the beam axis.
The innermost ITS layer is mounted on a beryllium beam pipe with an outer radius of 19~mm (28~mm in Run~2), contributing to the improved track pointing resolution.
The ITS is surrounded by the Time Projection Chamber (TPC)~\cite{ALICETPC:2020ann,CERN-LHCC-2013-020}, a large-volume cylindrical chamber with high-granularity readout, now equipped with Gas Electron Multiplier (GEM) detectors replacing the previously used Multi-Wire Proportional Chambers (MWPC). The TPC covers the region $\sim 85$ < \textit{r} < $\sim 247$~cm and -250 < \textit{z} < 250~cm in the radial and longitudinal directions, respectively.
Both ITS and TPC have track reconstruction capabilities within the TOF acceptance, with the ITS covering $|\eta| < 1.2$, and the TPC providing full azimuthal coverage up to $|\eta| < 0.9$.
The TPC also provides PID information through the measured specific energy loss (${\rm d}E/{\rm d}x$).
The Fast Interaction Trigger (FIT)~\cite{Antonioli:1603472} serves as a collision trigger, online and offline luminometer, and it provides an initial evaluation of the vertex position together with an estimation of the particle multiplicity at forward rapidity.
FIT consists of three subsystems composed in five standalone detector elements. They are asymmetrically located in forward regions on both sides from the interaction point.
The FIT subsystems provide redundancy for the measurements specified above, while each subsystem is optimized differently. The FIT Vertex-Zero (FV0) subsystem provides an estimate of collision centrality, the Forward Diffractive Detector (FDD) subsystem enables the measurement of collision cross sections, and the FIT Time-zero (FT0) subsystem provides a precise measurement of the collision time.
In particular, the FT0 subsystem is composed of two arrays of fast Cherenkov modules (FT0A and FT0C) on each side of the interaction point.
The radial distance along the beam axis between the interaction point and the FT0A module is 3.3~m, corresponding to a $3.5 < \eta < 4.9$ coverage, while for the FT0C module, the distance is 843~mm, covering $-2.1 < \eta < 3.3$~\cite{ALICE:2023udb}.

The TOF system covers the pseudorapidity interval $-0.9 < \eta < 0.9$ with full azimuthal acceptance.
The detector has a cylindrical symmetry and is located at an average distance of 3.8~m from the beam pipe, spanning an active area of 141~${\rm m}^{2}$.
The detector is made of 1593 Multigap Resistive Plate Chambers (MRPC) with a sensitive area of $7.4 \times 120~{\rm cm}^{2}$.
Each MRPC is segmented into 96 readout pads with an area of $2.5 \times 3.5~{\rm cm}^{2}$.
The TOF system measures the time of arrival of particles at the detector (\ttof).
Its time resolution drives the separation power between particle species~\cite{Antonioli:1603472}.
In addition, particle separation in the momentum interval of interest relies on a precise determination of the event collision time (\evtime), the track length ($l$), and the particle's momentum ($p$).
The distinction between different particle species is obtained via their mass hypothesis $m = \frac{|\overline{p}|}{c} \cdot \sqrt{c^{2}(\frac{\ttof - \evtime}{l})^{2} - 1}$.
The track length and the momentum measurement are obtained with the ITS and TPC detectors.
The \evtime is provided by the FT0 detector and by the TOF system itself~\cite{ALICE:2016ovj}.
The extensive upgrade of the TOF detector, which was realized during LS2, is described in Sec.~\ref{upgradetof}.

\subsection{Upgraded TOF and readout}
\label{upgradetof}

To address the challenges of the new data-taking conditions, it was required to upgrade the readout electronics of the TOF detector.
The goal was to accomplish a continuous readout, matching the readout scheme of the tracking detectors to allow particle identification in the intermediate momentum range for the full collected data sample.
The continuous readout was obtained with relatively limited interventions thanks to the small intrinsic dead time (\SI{\sim 10}{\nano\second}) of the MRPC detector and its front-end electronics, and making use of the buffering capabilities for digitized data available in the High Performance TDC (HPTDC) boards~\cite{Akindinov:2004gf}.
The HPTDC allows the readout trigger latency to be programmable over a large time range and supports overlapping trigger windows.
During Runs~1 and 2, due to the limited high-rate capability of the ALICE barrel detectors, the trigger rate was limited to a few kHz.
The internal HPTDC buffers for the TOF detector were configured with a latency window of \SI{7700}{\nano\second}, corresponding to the latency of the triggers reaching the TOF processing electronics.
The matching window was set to \SI{600}{\nano\second} to comfortably collect all hits, registered in the TOF detector, associated with the triggered collision.
However, with a programmable latency and matching windows, continuous readout may be achieved by applying a periodic trigger with a frequency $f_{\rm T}$ and a matching window $m_{\rm w} = 1/f_{\rm T}$.

\begin{figure}[!htbp]
  \centering
  \begin{minipage}{0.54\textwidth}
    \includegraphics[width=\textwidth]{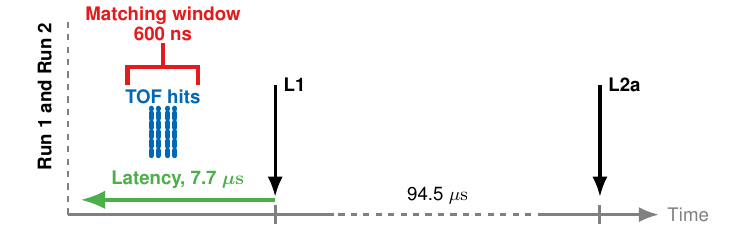}
    \includegraphics[width=\textwidth]{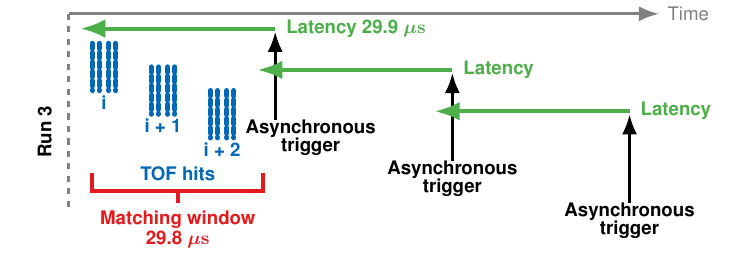}
  \end{minipage}
  \begin{minipage}{0.45\textwidth}
    \includegraphics[width=\textwidth]{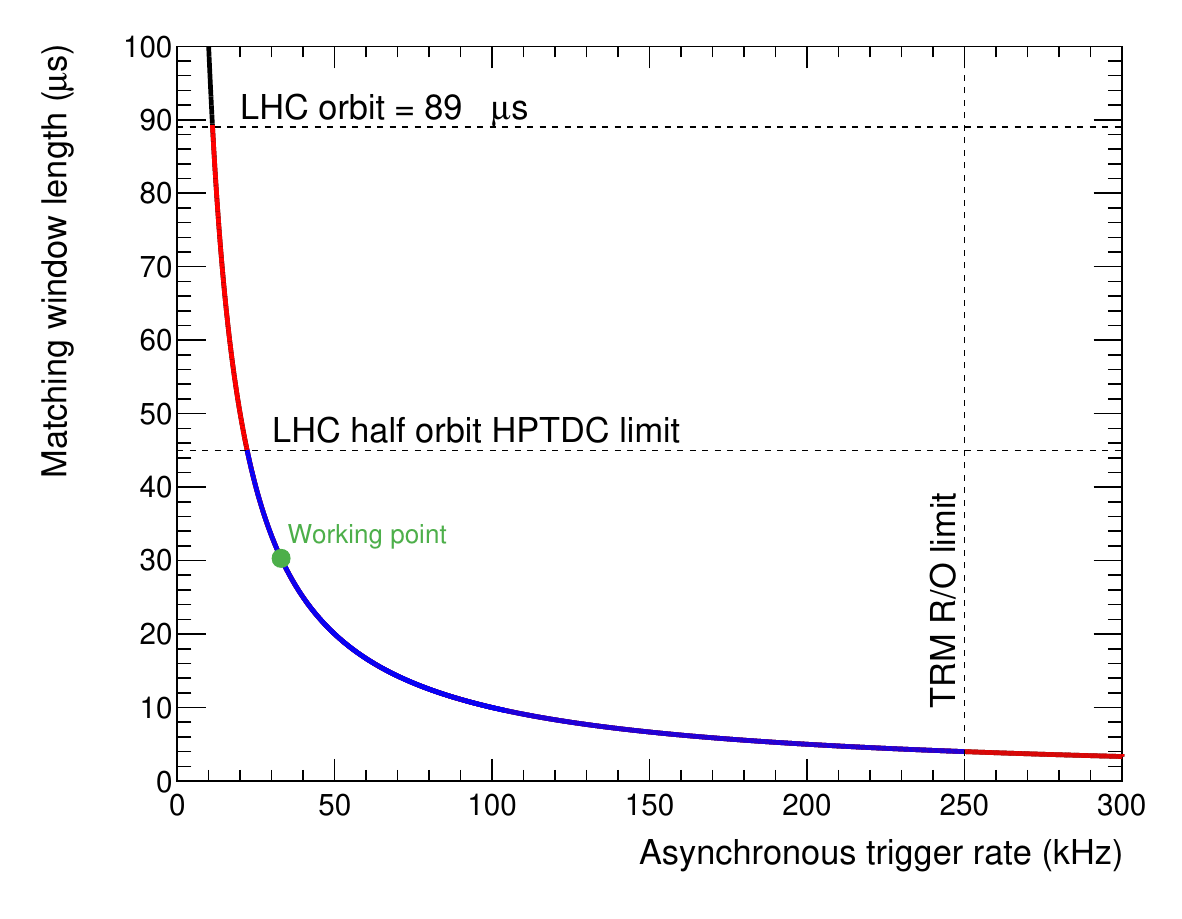}
  \end{minipage}
  \caption[TOF continuous read out implementation]{
  (left) HPTDC programming schema adopted during Run~1 and 2 (top left) and during Run~3 (bottom left).
  As described in detail in~\cite {ALICE:2023udb}, in Run~3, all triggers used previously are replaced by a periodic trigger (asynchronous with respect to actual collisions) at fixed bunch crossing number ($f=33.4$ kHz), mimicking a continuous readout and covering a complete LHC orbit within three triggers.
  All hits (short vertical blue lines) are read out and can be associated with physical events at a later stage.
  (right) Possible selection of parameters (asynchronous trigger frequency $f_\mathrm{T}$ and matching window width $m_\mathrm{w}$) to realize a continuous readout.
  The blue curve shows the region of allowed values.
  The red boundaries illustrate system limitations from the HPTDC matching algorithm and HPTDC readout time constraints.
  The green circle marks the chosen point of operation, with an asynchronous trigger rate of $\sim 33~{\rm kHz}$ and a matching window of $28.8~\mu s$.
  }
  \label{fig:cro}
\end{figure}

\Fig{fig:cro} (left) illustrates the underlying idea of the readout scheme, comparing the Run~3 approach to what was implemented during Run~1 and Run~2.
Before Run~3, the two trigger levels L1 and L2a received from the triggering hardware~\cite{ALICE:2014sbx} were used to read out the hits and match them to the triggered event, while in Run~3, all hits are read out by using an asynchronous trigger with a constant frequency.
\Fig{fig:cro} (right) shows the curve of allowed values (blue part of the curve), together with the limitations of the system: on the one hand, the latency window cannot be set at a value larger than half of an LHC orbit; on the other hand, as discussed in the ALICE Readout Upgrade TDR~\cite{Antonioli:1603472}, the trigger frequency cannot be too high, due to the HPTDC readout time inside the TDC readout module (TRM).
More generally, the readout time on average must be lower than $1/f_{\rm T}$.
After some optimization and a commissioning phase, an optimal operation point was found to be $f_{\rm T} = 3/t_{\rm LHC-orbit} \sim 33~{\rm kHz}$.

In order to keep up with the planned increase of the interaction rate (up to 1~MHz in pp collisions and 50~kHz in Pb--Pb collisions, after LS2), a new Digital Readout Module 2 (DRM2) was designed~\cite{Falchieri:2018fqw}.

The readout architecture of the upgraded ALICE system is based on the Common Readout Units (CRUs)~\cite{Buncic:2011297}, a standardized Peripheral Component Interconnect Express (PCIe) Field-Programmable Gate Array (FPGA)-based optical I/O processor module used by all upgraded detectors for data readout and configuration.
The detectors are configured via the detector control system (DCS), which is connected to the detector front-end electronics via the CRU.
The data produced by the TOF detector is handled by the Online \& Offline processing farm (\Otwo), consisting of first-level processors (FLPs) and event processing nodes (EPNs).
Depending on the detector implementation, the readout data are reformatted or compressed either in the front-ends, the CRUs, or in the FLPs~\cite{Buncic:2011297}.

\subsection{TOF calibrations in the continuous readout scheme}
Detector calibrations are performed at different stages of data taking and reconstruction to reach the best performance in terms of efficiency, noise rejection, and resolution, or, in simulations, to reproduce the conditions of the data taking.
The time granularity of the calibration objects is usually set at the level of a single run, or when conditions are expected to change rapidly (such as the LHC clock phase), in 5-minute sub-run time intervals.

Along the full chain of the data taking/reconstruction, there are three subsequent calibration stages:
\begin{enumerate}
  \item DCS calibration performed during the start of the run from the TOF slow control system;
  \item synchronous calibration done during the data taking on the EPN farm;
  \item asynchronous calibration required in the offline reconstruction.
\end{enumerate}

The DCS calibration entails the production and storage of the TOF active channel map at the beginning of the run.
An updated map is sent as soon as some channels are switched off during the run, together with the timestamp when the change occurred.
This object is then stored in the Condition and Calibration Data Base (CCDB).

The synchronous calibration is based on the synchronous reconstruction on EPNs, and provides a first version of time calibration objects. In addition, the frequency of decoding errors occurring during data taking is also monitored since it is needed for anchoring Monte Carlo simulations to real data conditions.
It is important to note that during synchronous processing, the precision of time calibration is limited due to reduced tracking performance — since the tracking detectors are also calibrated at this stage — and limited statistics. For this reason, refined calibrations are finally calculated asynchronously offline with a dedicated calibration pass performed on 10\% of the full statistics.
After the calibration reconstruction pass, TOF is declared ready for the reconstruction with the full statistics.

Time calibrations are divided into two main categories:
\begin{enumerate}
  \item TOF/ALICE clock shift relative to the LHC clock (done every 5 minutes),
  \item channel-by-channel calibrations (global offsets done run-by-run, charge dependence determined over several runs).
\end{enumerate}

\begin{figure}[!htbp]
  \centering
  \includegraphics[width=0.8\textwidth]{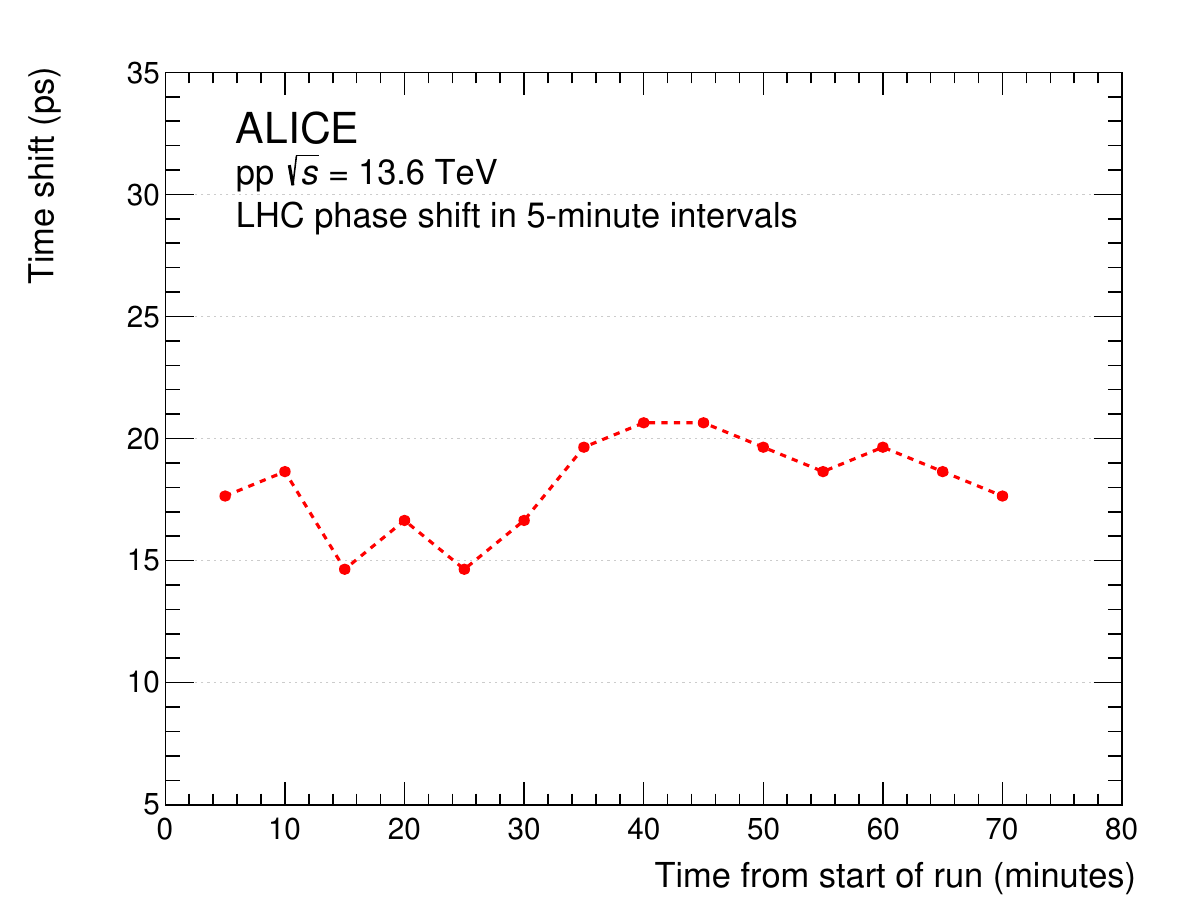}
  \caption{
    Example of the TOF clock alignment for a single run extracted in 5 minute intervals. The statistical error is significantly smaller than the marker size.
  }
  \label{fig:LHCphase}
\end{figure}

In \Fig{fig:LHCphase}, the trend of the TOF clock shift is reported as a function of time for one selected run; the analysis was performed by using five-minute time intervals.
Such a time interval is an acceptable compromise for following all relevant variations while keeping the number of calibration objects at a manageable level.

To reach the target timing resolution, channel-by-channel calibrations have to be measured independently for different values of the deposited charge in the active volume of the detector since the time needed before the signal rises above the threshold depends on it. Since the charge released in the MRPC cannot be measured directly, the time-over-threshold (ToT) is used as a proxy for it.
These calibrations are referred to as time-slewing corrections. Due to statistical limitations, only the global channel offsets can be calculated at the level of a single run while the time-slewing corrections are extracted by combining consecutive runs.

Independently of the granularity, all TOF time calibrations are calculated using the information from reconstructed tracks matched with a TOF cluster.
The values are obtained by fitting the Gaussian peak of the \deltaPi distribution, defined as the difference between the time measured by TOF (\ttof) and the expected time for a charged pion ($t_{\rm exp}(\pi)$), calculated from the particle's track length and momentum.
This procedure guarantees that the resulting \deltaPi distribution is centered around 0, once the residual offsets are subtracted.
The typical statistics required to constrain a single calibration parameter are of the order of 100 entries.
Since TOF has approximately 150k channels, about $5\cdot 10^{6}$ pp collisions are needed to calibrate the full detector.
In a typical Run 3 run (e.g. pp at 500~kHz) such statistics are reached after 15 minutes of data taking (assuming 1\% of the data to be reconstructed in the synchronous reconstruction).
Based on past experience, the time evolution of these corrections is expected to be very slow and independent of the reconstruction settings.

\begin{figure}[!htbp]
  \centering
  \includegraphics[width=0.75\textwidth]{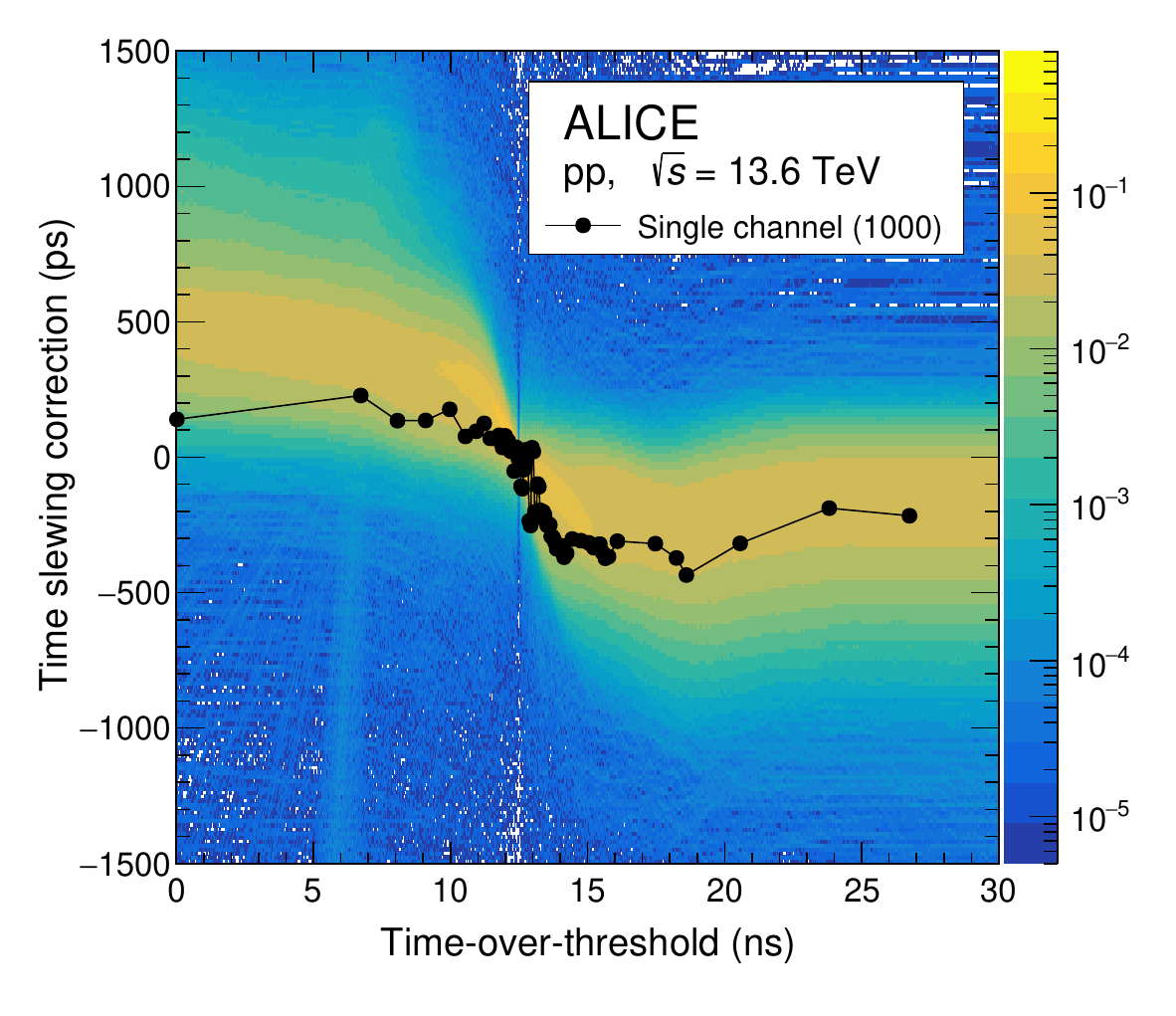}
  \caption{
    Distribution of time-slewing corrections: an example for one single channel is provided as a reference superimposed on a color-scale distribution corresponding to the density of all channel-calibration curves.
    A reference at zero was taken for all channels at ToT = 12.5 ns.
    The reported time-slewing calibration was obtained from the data taken with pp collisions at $\sqrt{s} = 13.6$~TeV. Sufficient statistics are required to ensure that the statistical uncertainty on the extracted correction remains below 10~ps.
    The shape of the correction is driven by the rise-time dependence of the discriminator threshold.
  }
  \label{fig:TimeSlewing}
\end{figure}

In \Fig{fig:TimeSlewing}, time-slewing corrections are shown for all TOF channels, with a single channel highlighted to illustrate the typical ToT granularity achieved with the 2022 data-taking statistics.
These are needed to compensate for the delay in the leading time due to the finite rise-time going above the threshold in the discriminator; otherwise, the rise-time dependence on the charge would introduce an extra smearing in the final TOF resolution.

\begin{figure}[!htbp]
  \centering
  \includegraphics[width=0.8\textwidth]{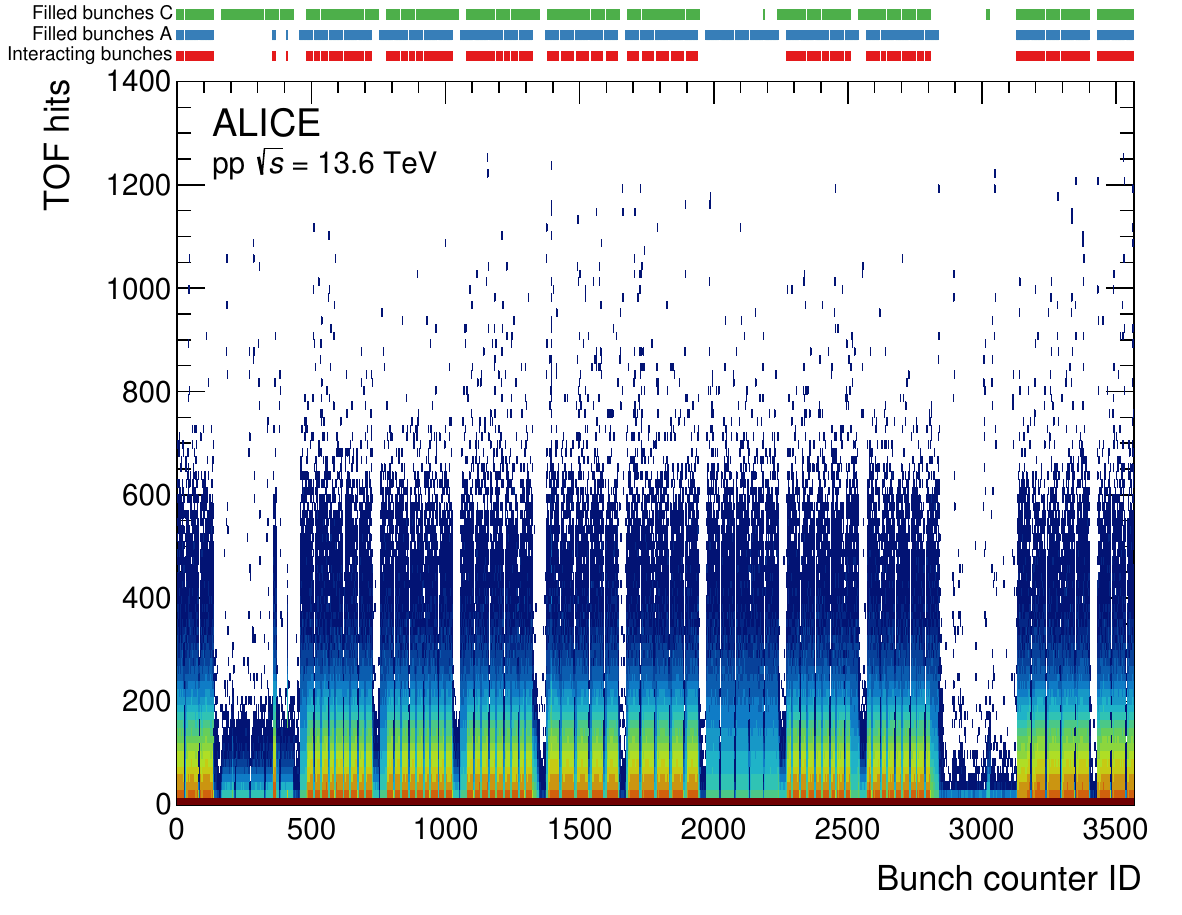}
  \caption{
    Hit multiplicity as a function of the bunch counter ID, blue colors represent lower counts.
    The structure of the LHC filling scheme for the data collected is reported for both beams (A and C) in the top section of the plot, the filling scheme is visible in the TOF hit multiplicity.
  }
  \label{fig:FillingScheme}
\end{figure}

The present TOF acquisition and calibration chain allows one to follow the event activity at the level of a single bunch crossing.
\Fig{fig:FillingScheme} shows the number of hits in the TOF as a function of the bunch counter ID, which uniquely identifies each potential bunch crossing.
This quantity is essential to verify that the noise levels of the detector are under control, as the majority of the hits have to occur around a bunch crossing with filled buckets (i.e., time slots in the accelerator where particle bunches can be placed).
The filling scheme of the LHC, corresponding to each bunch crossing, is reported for each of the beams in the top part of the plot.
The gaps show where particular bunch crossings do not have filled beam buckets and these are used to verify that the TOF detector noise levels are acceptable ($\approx 0.1~ \rm{Hz/cm^2}$).
This noise is a typical quantity that is monitored during data acquisition in the TOF Quality Control.

\section{Data analysis}\label{sec:pap:analysis}

This analysis was carried out on data collected in 2022 with pp collisions at $\sqrt{s} = 13.6$~TeV.
Tracks reconstructed with the ITS and TPC are propagated to the TOF detector for the matching algorithm to assign a value for the particle time-of-flight.
Only tracks with a TOF match are considered.
The expected time of flight of a track (\texp) is computed during track propagation from the primary vertex to the TOF-matched hit for each mass hypothesis.
About $10^{9}$ events collected at an average interaction rate of about 500~kHz were used.
In the continuous readout scenario, where events are not hardware-triggered, the definition of an event coincides with the collision vertex of reconstructed tracks.
To ensure that the reconstructed collision comes from a genuine pp event, the timing information from the two FT0 arrays must match the timing of the LHC bunch crossing.
High-quality tracks are selected by applying the following criteria separately on each track.
Reconstruction requires the use of both the ITS and the TPC information.
A minimum of four ITS hits is needed, with at least two located in the first three layers.
Tracks with fewer than 70 out of 152 crossed rows in the TPC are discarded.
To further ensure quality, the reduced chi-squared of the track fit quality, defined as $\chi^2/{\rm NDF}$, where $\rm NDF$ is the number of degrees of freedom, is required to be lower than 2.

In addition, primary tracks, i.e. tracks that point to the primary vertex of the collision, are selected to avoid the contribution of delayed signals from particles originating from displaced secondary vertices.
These are discarded by requiring the distance of closest approach of the track to the primary vertex to be smaller than seven times its expected resolution.
A dedicated study using a Monte Carlo simulation was carried out to optimize these selection criteria as described in~\cite{ALICE:2025cjn}.

\begin{figure}
  \centering
  \includegraphics[width=0.6\textwidth]{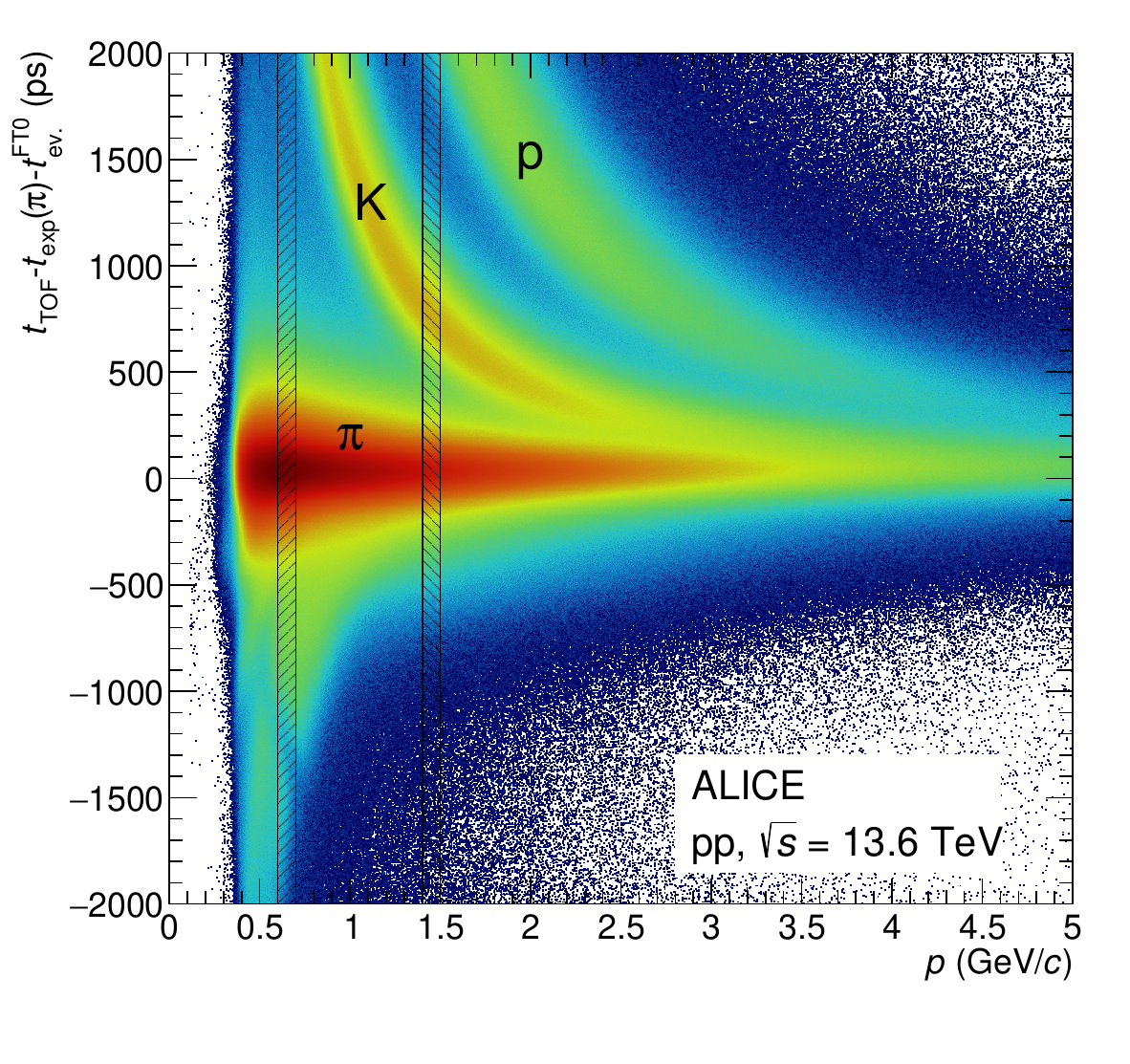}
  \caption{
    Time difference measured with TOF with respect to the expected value for the charged-pion mass hypothesis as a function of momentum, blue colors represent lower counts.
    The two vertical shaded regions indicate momentum ranges used to obtain the final time resolution for the two methods presented in the text.
  }
  \label{fig:SeparationTOF}
\end{figure}

The separation achieved for the different particle species with the TOF detector is shown as a function of momentum ($p$) in \Fig{fig:SeparationTOF}.
The separation is computed under the mass hypothesis of the charged pions.
The kaon and proton bands are clearly visible and separated from the pion band at low momenta.
The merging of the bands at high momentum is driven by the timing resolution.
The better the resolution, the higher the momentum limit for the particle identification becomes; thus, considerable effort is required to improve the resolution as much as possible.

The detector resolution is measured using two methods.
First, we provide a self-consistent measurement by using the variable \doubledelta, defined as the difference between the time measured for two tracks matched with TOF in the same event.
This strategy has the advantage of being independent of the definition of the collision time.
Secondly, the time resolution was measured by calculating the quantity \deltaGeneralTZAC, where \evTimeTZAC is the event time obtained with the FT0 detector.
It has to be noted that the TOF signal shape is not entirely Gaussian, but features a non-Gaussian tail.
  This effect was marginally observed also when the detector operated under ideal conditions (i.e., during beam tests), and its underlying cause is not fully understood.
  However, it remained relatively stable throughout the data-taking campaigns of Run~1 and Run~2.
  To account for such non-Gaussian effects, and in agreement with previous works~\cite{ALICE:2014sbx, ALICE:2019hno}, an exponentially modified Gaussian distribution (EMG, \Eq{expotail}), i.e., a Gaussian distribution with an exponential tail on the right side of its peak, is used to extract the detector timing resolution.
  \begin{equation} \label{expotail}
    f(x) \propto
    \begin{cases}
      \frac{1}{\sigma \sqrt{2\pi}}
      e^{-\frac{1}{2} \left(\frac{x-\mu}{\sigma}\right)^2}
       & \mbox{if } x \leq \mu + \tau \\
      \frac{1}{\sigma \sqrt{2\pi}}
      e^{-\frac{1}{2} \left(\frac{\mu + \tau -\mu}{\sigma}\right)^2 -\tau \cdot \frac{x - \mu - \tau}{\sigma^2}}
       & \mbox{if } x > \mu + \tau
    \end{cases}
  \end{equation}

  With the detector properly calibrated, the tail becomes visible above approximately 1 $\sigma$ and is well-described by the EMG parameterization.
By comparison with a pure Gaussian distribution, it can be therefore observed that while for the Gaussian the probability above 2 $\sigma$ and 3 $\sigma$ is at the level of 2.3\% and 0.13\%, respectively, in a realistic case, as shown later, the fraction of signal above 2 $\sigma$ and 3 $\sigma$ slightly increases to 7\% and 2.3\%, reflecting, as expected, the larger contribution due to the tail.
  The reduction in efficiency when applying a 3 $\sigma$ cut is 3\%, and this effect is well reproduced by the Monte Carlo simulation.

    Considering that the non-Gaussian effects are present on the right side of the \deltaGeneralTZAC distribution, when applying the \doubledelta technique, one has to account for a reflected exponential contribution on the left side of the Gaussian peak.
    This is achieved by considering the fitting process, which utilizes a double exponentially modified Gaussian distribution (Double EMG, \Eq{expotail2}).
    \begin{equation} \label{expotail2}
      f(x) \propto
      \begin{cases}
        \frac{1}{\sigma \sqrt{2\pi}}
        e^{-\frac{1}{2} \left( \frac{x - \mu}{\sigma} \right)^2}
         & \mbox{if } \mu - \tau \leq x \leq \mu + \tau \\[10pt]
        \frac{1}{\sigma \sqrt{2\pi}}
        e^{-\frac{1}{2} \left( \frac{\tau}{\sigma} \right)^2 - \frac{\tau}{\sigma^2} (x - \mu - \tau)}
         & \mbox{if } x > \mu + \tau                    \\[10pt]
        \frac{1}{\sigma \sqrt{2\pi}}
        e^{-\frac{1}{2} \left( \frac{\tau}{\sigma} \right)^2 - \frac{\tau}{\sigma^2} (\mu - \tau - x)}
         & \mbox{if } x < \mu - \tau
      \end{cases}
    \end{equation}

\subsection{Resolution measurement via the two-track difference}
The first method to obtain the time resolution of the TOF considers a self-consistent measurement based on a two-track difference.
A double-delta variable between two tracks of the same event, using a given particle mass hypothesis, here for the pion, is defined as:
\begin{equation}
  \label{ref_doubledelta}
  \doubledelta = [\deltaTwo - \evTime] - [\deltaOne - \evTime] = \doubledeltaLong
\end{equation}
The advantage of using $\doubledelta$ is that it does not require any knowledge of the event collision time, which varies event-by-event depending on the longitudinal spread of the colliding bunches.

The approach is to consider one track (1) as a ``reference'' in a well defined kinematic region ($600 < p < \mevc{700}$, see the left vertical band in \Fig{fig:SeparationTOF}) and the second one (2) as a probe at any given momentum ($p$).

The resolution of $\doubledelta$ is defined by two contributions, the resolution of the reference and that of the probe, as:

\begin{equation}
  \label{sigma_doubledelta}
  \sigmadoubledelta = \sigma(p) \oplus \sigmaRef
\end{equation}

Since the  contributions of the two tracks are independent and all sources of uncertainty are well approximated by Gaussian distributions, they all result in a sum in quadrature, indicated by $\oplus$. As with \Eq{ref_doubledelta}, each term contains a contribution to the resolution which comes from the time-of-flight measurement ($\sigma_{\rm TOF}$) and one from the momentum-dependent expected time from tracking ($\sigma_{\rm t-exp}(p)$):
\begin{equation}
  \sigma(p) = \sigma_{\rm TOF} \oplus \sigma_{\rm t-exp}(p)
\end{equation}

It is worth noting that $\sigma_{\rm t-exp}(p)$ strongly depends on momentum and on the mass of the particle since $t_{\rm exp}$ inherits all the uncertainties, during track propagation, about the particle velocity.
Due to the energy loss along the track trajectory, such uncertainties are larger for lower beta (lower momenta) and at a given momentum for heavier particles.

By choosing the particles under study in the same kinematic region as the reference particles ($\sigma(p)=\sigmaRef$) it is possible to measure $\sigmaRef$:
\begin{equation}
  \sigmadoubledeltaref = \sigmaRef \oplus \sigmaRef = \sqrt{2} \cdot \sigmaRef
\end{equation}
\begin{equation}
  \sigmaRef = \sigmadoubledeltaref / \sqrt{2}
\end{equation}

Once $\sigma_{\rm reference}$ is known, the analysis can be repeated by varying the momentum of the particles under study to extract, using \Eq{sigma_doubledelta}, the single particle resolution
\begin{equation} \label{eqsub}
  \sigma(p) = \sigma_{\rm TOF} \oplus \sigma_{\rm t-exp}(p) = \sqrt{\sigma_{\Delta\Delta}^{2}(p) - \sigmaRef^{2}}
\end{equation}

As it will be shown in Sect.~\ref{sec:pap:results}, when the particle momentum is high enough (e.g, for pions $p > \gevc{1}$), the contribution from $\sigma_{\rm t-exp}$ from tracking becomes negligible, and only the TOF component contributes to the resolution.
In addition, in order to work under the assumption of a negligible tracking contribution, tracks at $|\eta| < 0.5$ are selected; this requirement matches common selection criteria employed in analysis (e.g., in~\cite{ALICE:2019hno}).

\begin{figure}
  \centering
  \includegraphics[width=0.6\textwidth]{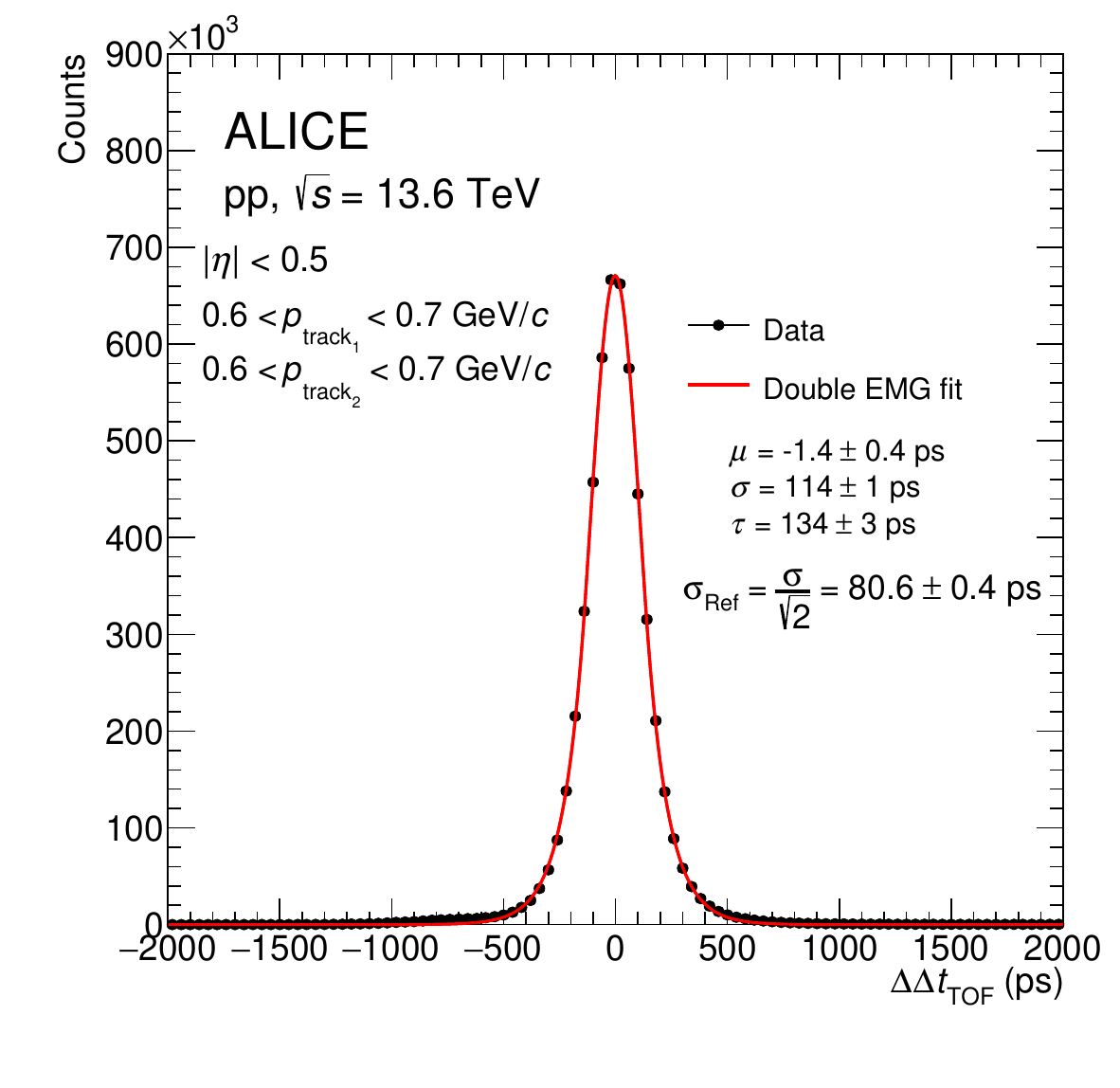}
  \caption{
    \doubledelta distribution for the reference tracks, selecting both \trkOne and \trkTwo with a momentum in the range $ 600 < p < \mevc{700}$, represented by the black-shaded band in \Fig{fig:SeparationTOF}. The fit is done over the full range of the histogram.
  }
  \label{fig:DeltaDeltaRef}
\end{figure}

In this analysis, the ``reference resolution'' \sigmaRef  is extracted when both \trkOne and \trkTwo have $600 < p < \mevc{700}$.
The \doubledelta distribution of the selected tracks, together with the mean $\mu$ and sigma $\sigma$ of the fitted Gaussian distribution and the resulting resolution \sigmaRef are reported in \Fig{fig:DeltaDeltaRef}.

For the data considered, the obtained reference resolution is $\sigmaRef = \DoubleDeltaRefSigmaValue \pm \DoubleDeltaRefSigmaError~{\rm ps}$.
As mentioned, the procedure is repeated for various values of the momentum, according to \Eq{eqsub}, and  using \sigmaRef to calculate $\sigma(p)$ as a function of momentum.

\subsection{Resolution extracted using the event collision time}

The second method used to evaluate the TOF resolution employs a more ``classical'' approach, similar to what was done previously in~\cite{Akindinov:2013tea}.
  The measurement of the separation between particle species \deltaGeneral uses event collision times obtained from the FT0 detector.
  The event collision time is provided by combining the timing information from the FT0A and FT0C modules.

  \begin{equation}
    \label{ev_time_ft0}
    \evTimeTZAC = (\evTimeTZA + \evTimeTZC) / 2
  \end{equation}
  The delay due to the distance from the primary vertex cancels out in \Eq{ev_time_ft0}, providing an estimate of the event collision time, which is independent of the TOF measurements.
  The contribution of the \evTimeTZAC to the \deltaGeneralTZAC can then be then written as:

  \begin{equation}
    \label{sigma_t0ac}
    \sigma_{\deltaGeneralTZAC} = \sigma_{\rm TOF} \oplus \sigma_{\rm t-exp}(p) \oplus \sigma_{\evTimeTZAC}
  \end{equation}

  Assuming that at high momenta the uncertainty related to the tracking $\sigma_{\rm t-exp}(p)$ becomes negligible

  \begin{equation}
    \label{sigma_t0ac_highpt}
    \sigma_{\deltaGeneralTZAC} = \sigma_{\rm TOF} \oplus \sigma_{\evTimeTZAC}
  \end{equation}

\begin{figure}
      \centering
      \includegraphics[width=0.6\textwidth]{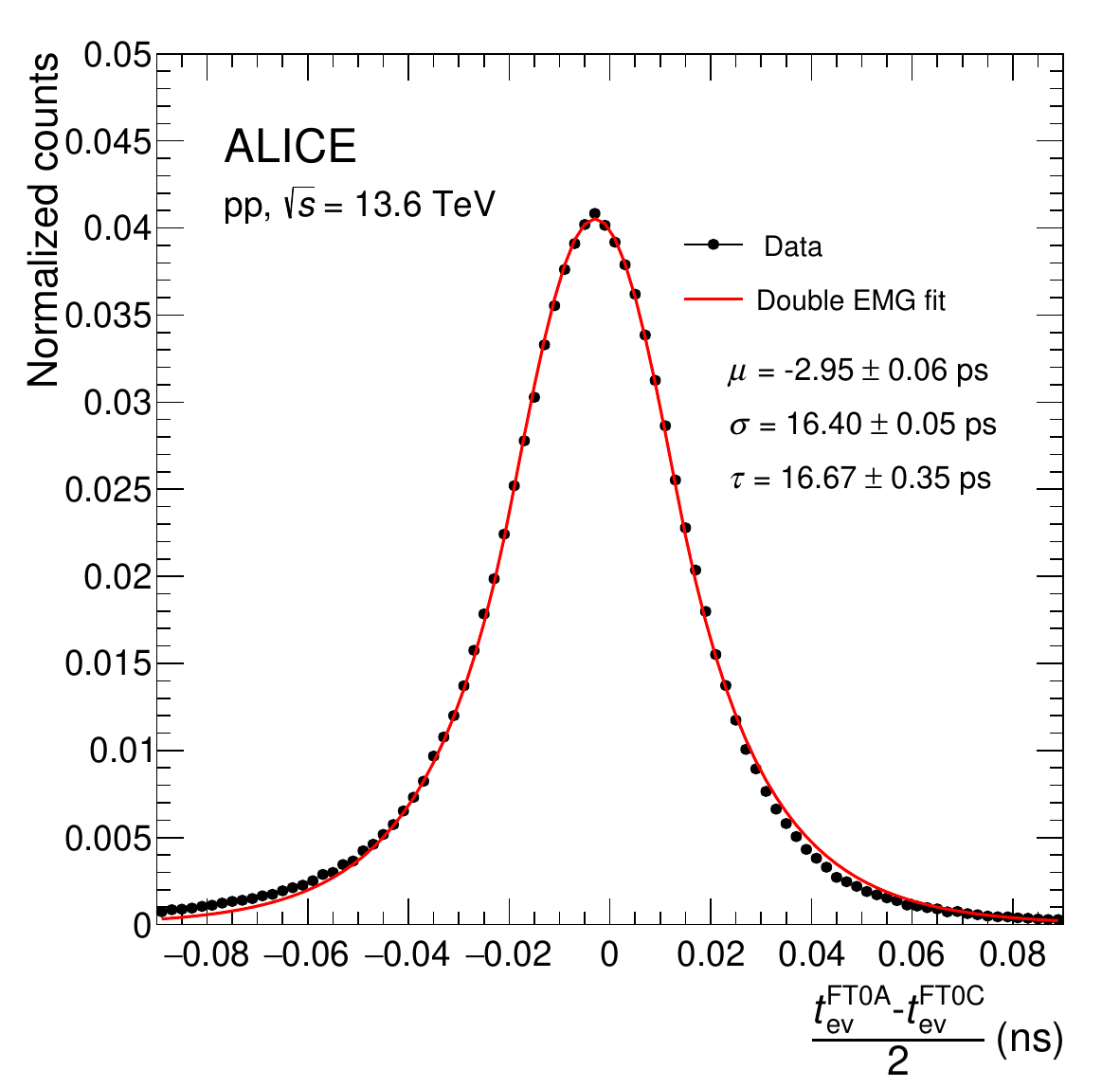}
      \caption{
        Event time resolution measured with the FT0 detector in pp collisions at $\sqrt{s} = \gevc{13.6}$, to account for non-Gaussian tails, a Double EMG function is used for the fit done over the full range of the histogram.
        The distribution is normalized to its integral.
      }
      \label{fig:EvTimeReso}
    \end{figure}

  The separation between different particle species obtained by using the FT0 information is shown in \Fig{fig:SeparationTOF}.
  The resolution is extracted at high momenta ($1.4 < p < \gevc{1.5}$, indicated from the black-shaded band in \Fig{fig:SeparationTOF}), where the contribution due to tracking is expected to be negligible.
  The momentum interval is chosen so that the peaks for pions and kaons are well separated.
    The resolution on the event time is obtained by considering the FT0A and FT0C, corrected for the primary vertex position, individually and it is reported in \Fig{fig:EvTimeReso}.
    In deriving the final result, the contribution of the FT0, which provides an event time resolution of approximately \FTZResoValue can be considered negligible once added in quadrature.

\section{Results}\label{sec:pap:results}

\begin{figure}[h]
  \centering
  \includegraphics[width=0.60\textwidth]{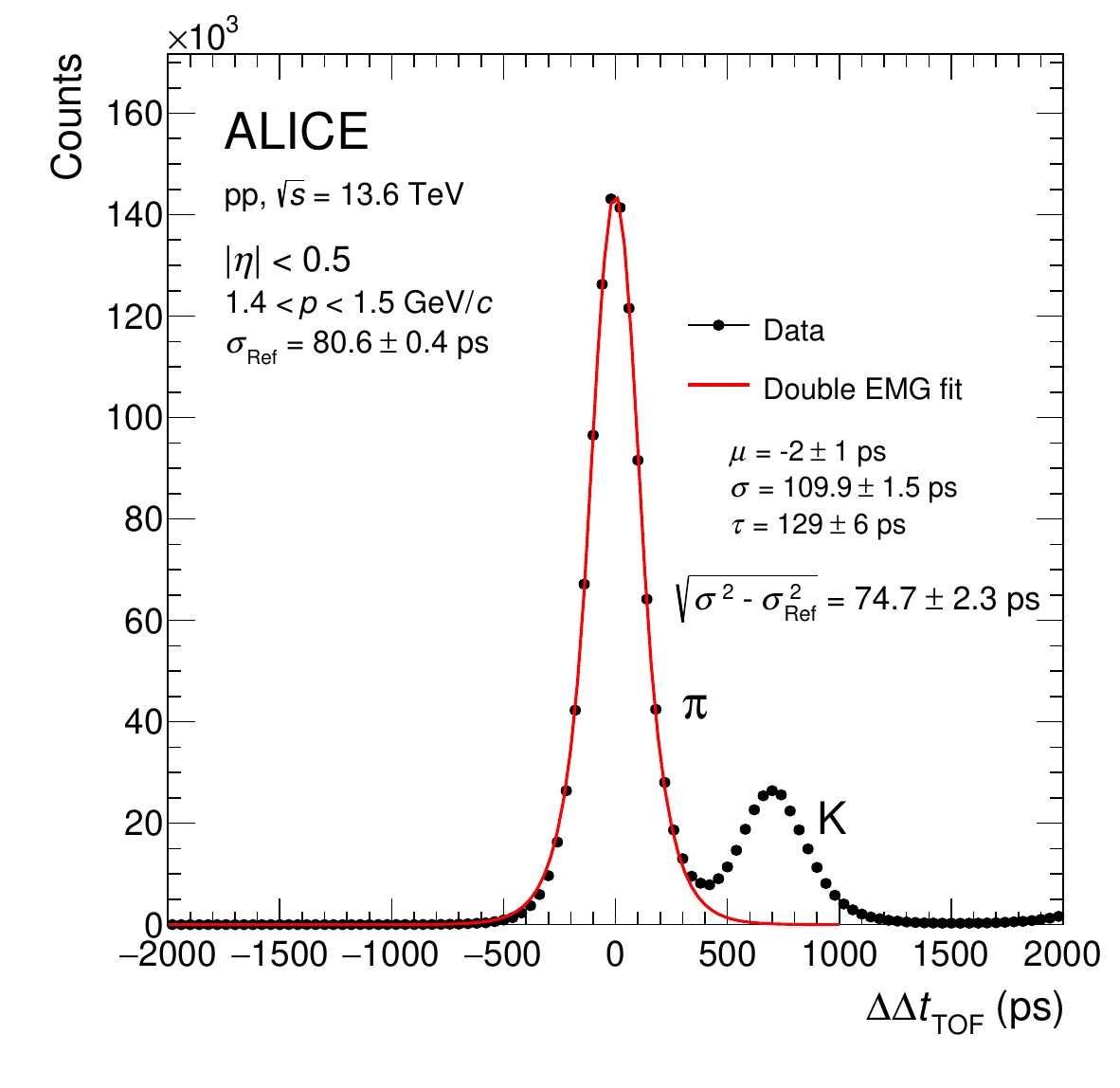}
  \caption{
    Distribution of \doubledelta obtained by considering the track under study in the momentum range $ 1.4 < p < \gevc{1.5}$.
    In this momentum range, the peaks corresponding to pions and kaons can be clearly distinguished.
    The red curve represents the Double EMG fit used to extract the \doubledelta time resolution.
    The fit is done over the full range, and the results obtained for different ranges are taken into account in the error.
  }
  \label{fig:ct_texp_DDvsP}
\end{figure}

\begin{figure}[h]
  \centering
  \includegraphics[width=0.60\textwidth]{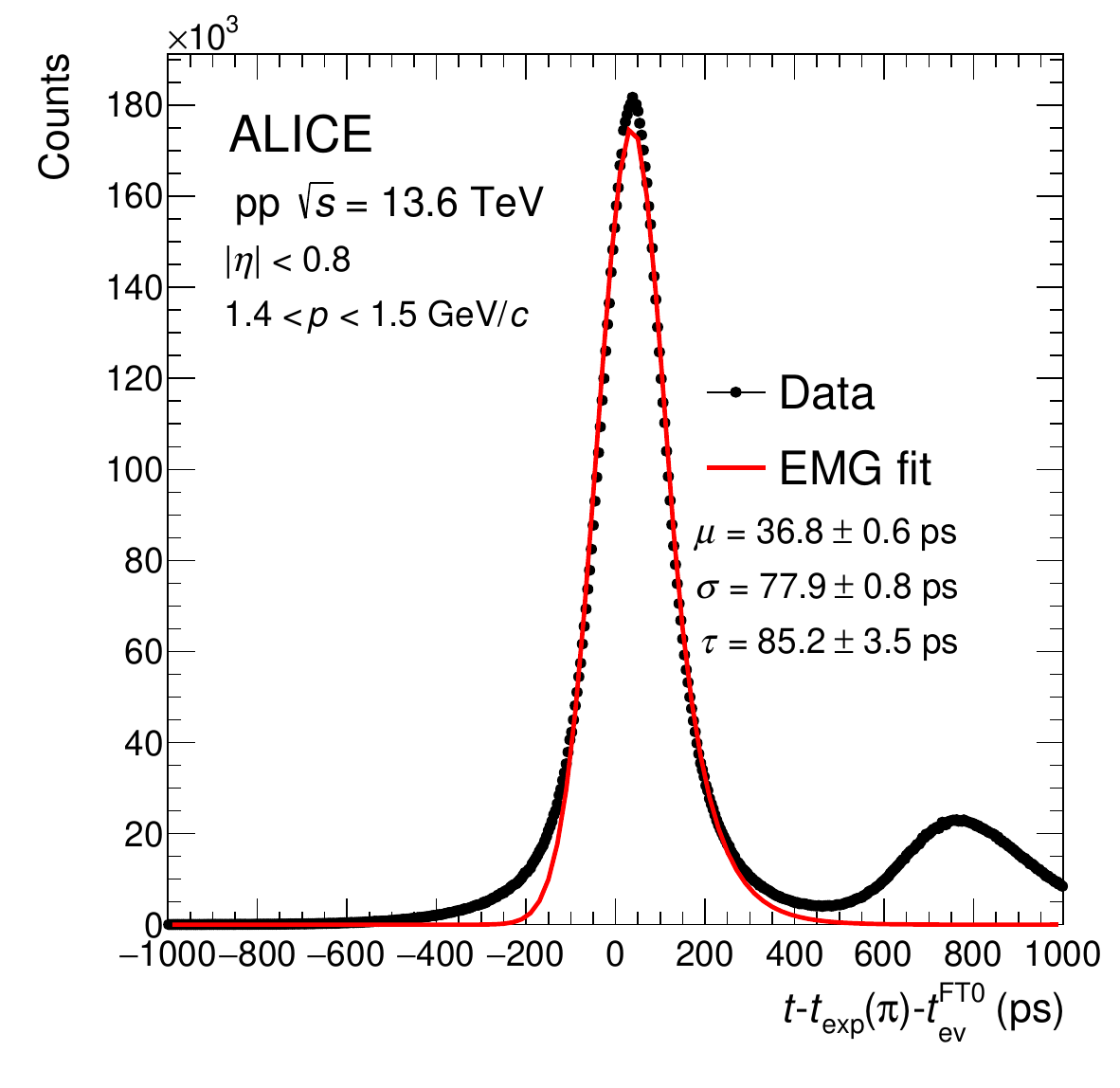}
  \caption{
    Distribution of the \deltaPiTZAC in the momentum range $ 1.4 < p < \gevc{1.5}$.
    The red curve represents the EMG fit used to extract the time resolution.
    The fit is done in the range (-200, 200) ps, and the results obtained from different ranges are taken into account in the error.
  }
  \label{fig:ct_texp_FT0vsP}
\end{figure}

The overall results obtained for the TOF timing resolution with the two methods introduced in Sect.~\ref{sec:pap:analysis} are presented here.
For both methods, the best case is defined for the particle under study having a momentum range $1.4 < p < \gevc{1.5}$. As it can be seen in \Fig{fig:SeparationTOF}, in this momentum range (denoted by the black region), the two peaks for pions and kaons are well separated. This kinematic region is chosen because the contribution of tracking is almost negligible and it is not possible to go to higher momenta where the two peaks are no longer resolved, without avoiding contamination from kaons.
The resulting \doubledelta distribution is shown in \Fig{fig:ct_texp_DDvsP}.
After the contribution of the resolution on the reference (extracted from \Fig{fig:DeltaDeltaRef}) is subtracted, using \Eq{eqsub}, the TOF resolution is found to be \DoubleDeltaResult.
The resolution extracted for $1.4 < \mom < \gevc{1.5}$ using the FT0 detector is reported in \Fig{fig:ct_texp_FT0vsP}.
This technique results in a TOF resolution of \StdResult in the $|\eta| < 0.5$ range, and \StdResultFullEta when averaged over the full $\eta$ range.

\begin{figure}
    \centering
    \includegraphics[width=0.6\textwidth]{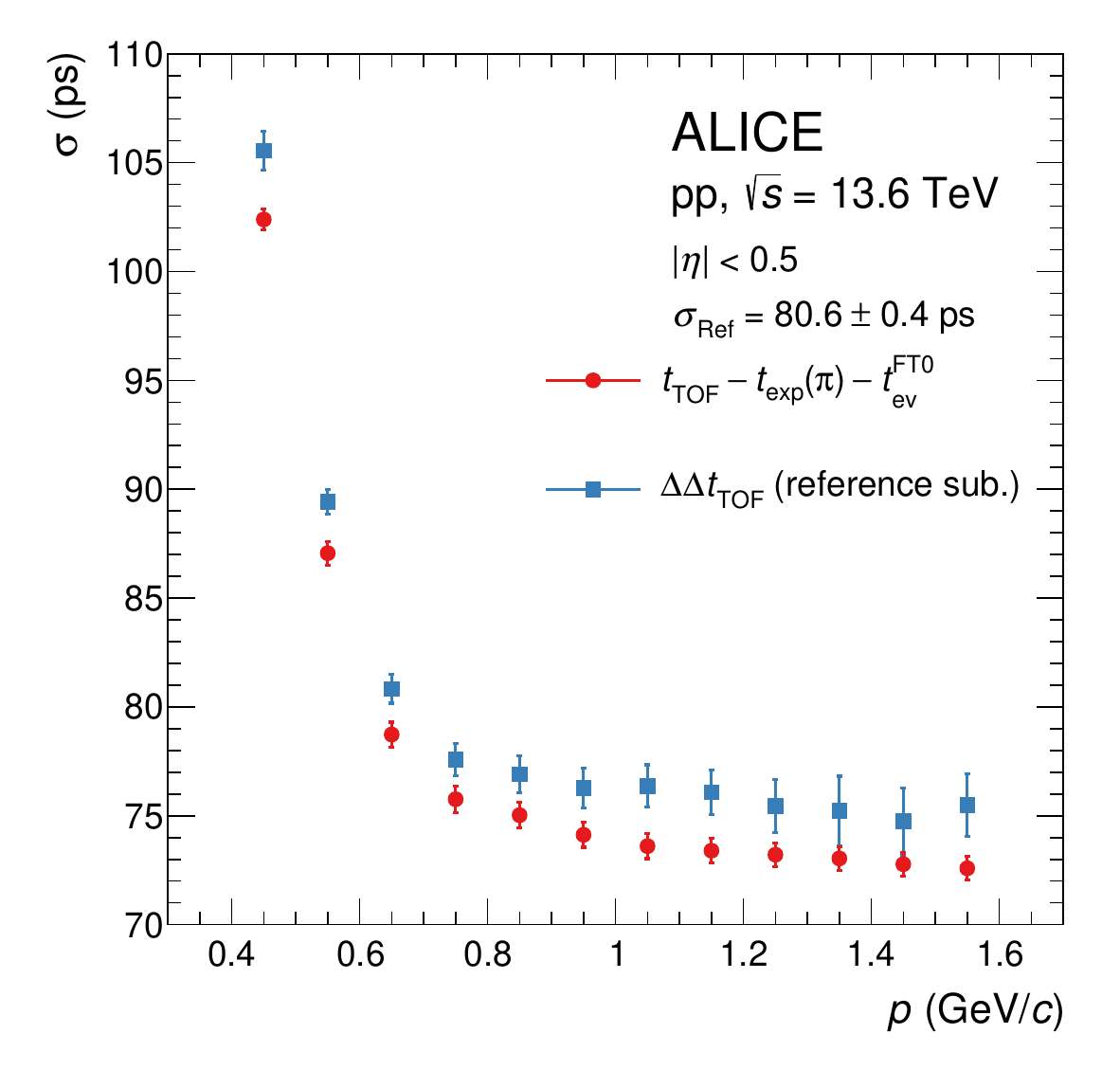}
    \caption{
      Timing resolution as a function of momentum \mom obtained using the \doubledelta method (\sigmaRef is already subtracted) and \deltaPiTZAC method.
      As the tracking momentum resolution improves, with increasing momentum, so does the timing resolution.
    }
    \label{fig:ResVSpt}
  \end{figure}

  The contribution of tracking to the TOF resolution ($\sigma(p)$) can be inferred using the \doubledelta method and by considering \deltaPiTZAC, as shown in \Fig{fig:ResVSpt}.
The resolution improves significantly -- from approximately 100 ps at a momentum of around \mom $\sim \mevc{500}$ — to better than 80 ps for the \doubledelta method and \StdResultValue~ps when using \deltaPiTZAC at higher momenta, where the tracking contribution becomes nearly negligible.

In conclusion, both methods yield consistent results, with an overall resolution of the TOF detector of better than 80 ps.

\begin{figure}
  \centering
  \includegraphics[width=0.49\textwidth]{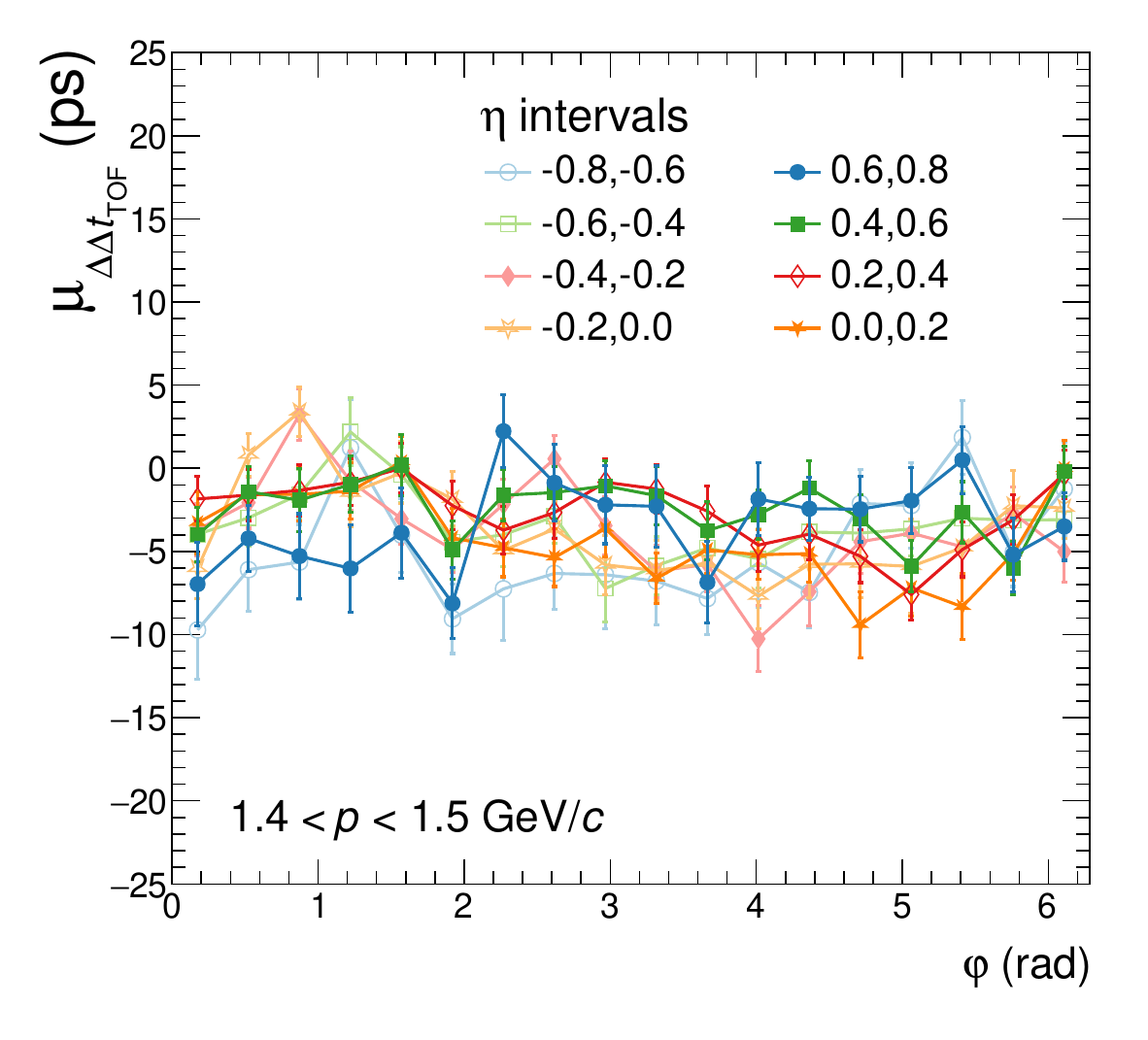}
  \includegraphics[width=0.49\textwidth]{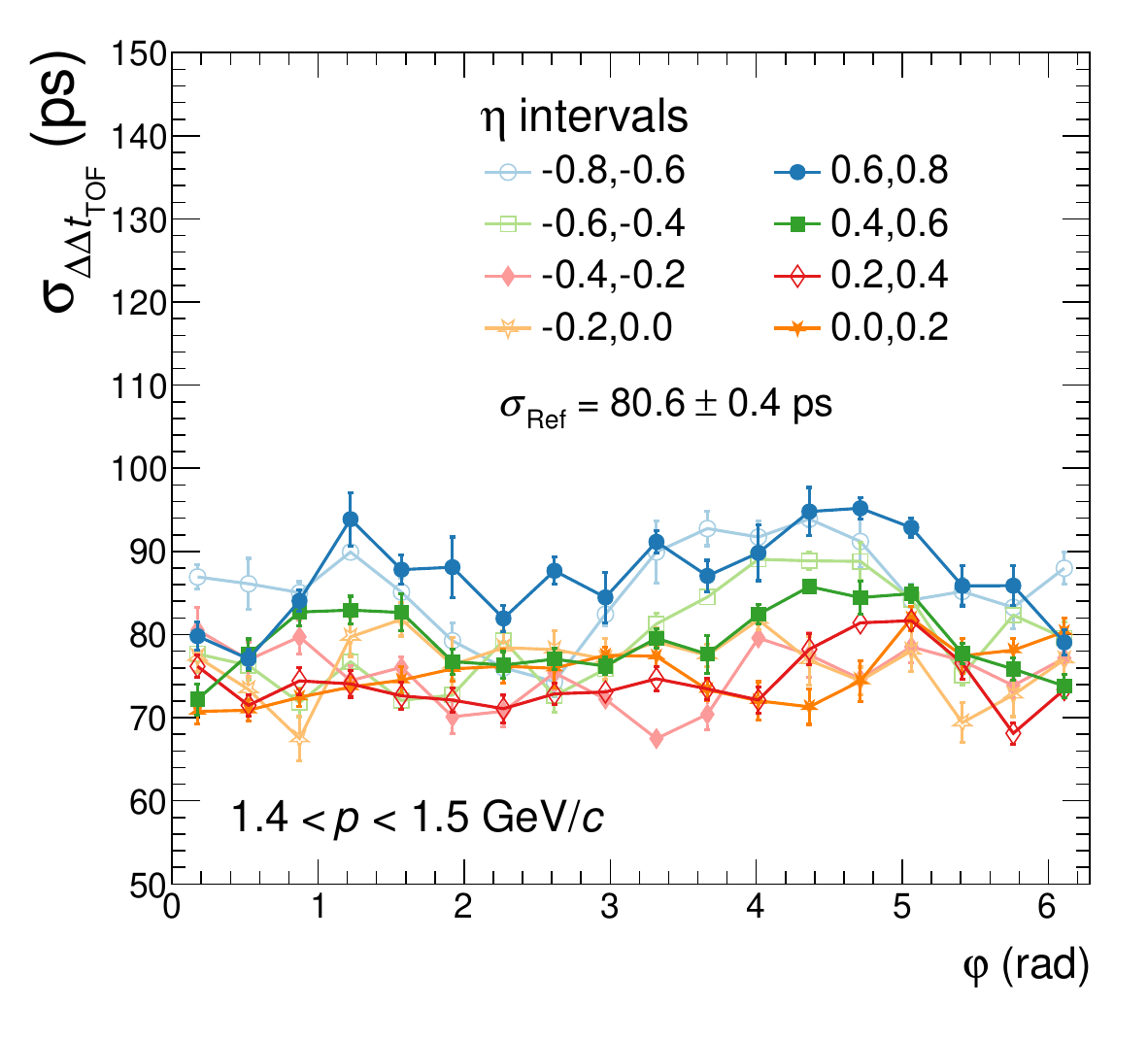}
  \caption{
    Distributions of the mean value (left) and sigma (right) using the \doubledelta method in the momentum range $ 1.4 < p < \gevc{1.5}$ as a function of $\varphi$ for different $\eta$ intervals.
  }
  \label{fig:ResoVsPhiEta}
\end{figure}

In order to verify the stability of the procedure, the analysis with \doubledelta was also repeated in different $\eta$-$\varphi$ slices within the same momentum interval ($1.4~<~p~<~\gevc{1.5}$).
The results for the mean position of the peak and resolution are reported in \Fig{fig:ResoVsPhiEta}.
The values of the mean are independent of both variables $\eta$ and $\varphi$, as expected, following the calibration procedure.
Similarly, the values of the resolution are independent of $\varphi$, while it appears to be slightly worse at larger $\eta$, on both the positive and negative sides.
This effect can be ascribed to residual imperfections in track quality for those regions.
It is worth noting that the reconstruction software undergoes a few changes between the calibration pass and the physics pass; these can lead to small deviations from 0 of the mean values.
In any case, all variations proved to be below 5 ps.
Indeed, the quality of the TOF calibrations is strongly coupled to the tracking performance, which is expected to improve over time during ALICE operations.
The differential analysis we performed indicates that there are still margins for improvement once the entire ALICE Run~3 procedure is better understood, after which a similar timing performance to that of 56~ps reached in Pb--Pb collisions at the end of Run~2 is expected~\cite{inproceedings}.

\section{Conclusions}\label{sec:pap:conclusions}

The PID capabilities of the ALICE experiment are essential for realizing its physics program.
The ALICE Time-Of-Flight detector is crucial to provide particle separation in the intermediate momentum region (0.5 < $p$ < \gevc{4}).
The recent upgrade of the TOF detector was mainly centered around its readout electronics to fit into the new data acquisition system.
A refurbishment of other TOF components was also carried out based on the experience gained in Run~1 and Run~2 to improve their operations.
As a result, the new offline handling of the data, calibration, and quality control, which were also developed, are presented in this paper.
The TOF PID performance is evaluated by measuring the detector resolution of the time-of-flight measurement using two methods: one purely relying on the TOF and tracking information presented here for the first time, and one that depends on the global event time determination.
The two methods yield consistent results, with an overall resolution better than 80~ps, reaching the target set out in the technical design report for the TOF system~\cite{CERN-LHCC-2000-012}.
The achievements presented in this paper confirm the capability of ALICE to maintain a reliable PID in the continuous readout mode of Run~3.


\newenvironment{acknowledgement}{\relax}{\relax}
\begin{acknowledgement}
\section*{Acknowledgements}

The ALICE Collaboration would like to thank all its engineers and technicians for their invaluable contributions to the construction of the experiment and the CERN accelerator teams for the outstanding performance of the LHC complex.
The ALICE Collaboration gratefully acknowledges the resources and support provided by all Grid centres and the Worldwide LHC Computing Grid (WLCG) collaboration.
The ALICE Collaboration acknowledges the following funding agencies for their support in building and running the ALICE detector:
A. I. Alikhanyan National Science Laboratory (Yerevan Physics Institute) Foundation (ANSL), State Committee of Science and World Federation of Scientists (WFS), Armenia;
Austrian Academy of Sciences, Austrian Science Fund (FWF): [M 2467-N36] and Nationalstiftung f\"{u}r Forschung, Technologie und Entwicklung, Austria;
Ministry of Communications and High Technologies, National Nuclear Research Center, Azerbaijan;
Rede Nacional de Física de Altas Energias (Renafae), Financiadora de Estudos e Projetos (Finep), Funda\c{c}\~{a}o de Amparo \`{a} Pesquisa do Estado de S\~{a}o Paulo (FAPESP) and The Sao Paulo Research Foundation  (FAPESP), Brazil;
Bulgarian Ministry of Education and Science, within the National Roadmap for Research Infrastructures 2020-2027 (object CERN), Bulgaria;
Ministry of Education of China (MOEC) , Ministry of Science \& Technology of China (MSTC) and National Natural Science Foundation of China (NSFC), China;
Ministry of Science and Education and Croatian Science Foundation, Croatia;
Centro de Aplicaciones Tecnol\'{o}gicas y Desarrollo Nuclear (CEADEN), Cubaenerg\'{\i}a, Cuba;
Ministry of Education, Youth and Sports of the Czech Republic, Czech Republic;
The Danish Council for Independent Research | Natural Sciences, the VILLUM FONDEN and Danish National Research Foundation (DNRF), Denmark;
Helsinki Institute of Physics (HIP), Finland;
Commissariat \`{a} l'Energie Atomique (CEA) and Institut National de Physique Nucl\'{e}aire et de Physique des Particules (IN2P3) and Centre National de la Recherche Scientifique (CNRS), France;
Bundesministerium f\"{u}r Forschung, Technologie und Raumfahrt (BMFTR) and GSI Helmholtzzentrum f\"{u}r Schwerionenforschung GmbH, Germany;
General Secretariat for Research and Technology, Ministry of Education, Research and Religions, Greece;
National Research, Development and Innovation Office, Hungary;
Department of Atomic Energy Government of India (DAE), Department of Science and Technology, Government of India (DST), University Grants Commission, Government of India (UGC) and Council of Scientific and Industrial Research (CSIR), India;
National Research and Innovation Agency - BRIN, Indonesia;
Istituto Nazionale di Fisica Nucleare (INFN), Italy;
Japanese Ministry of Education, Culture, Sports, Science and Technology (MEXT) and Japan Society for the Promotion of Science (JSPS) KAKENHI, Japan;
Consejo Nacional de Ciencia (CONACYT) y Tecnolog\'{i}a, through Fondo de Cooperaci\'{o}n Internacional en Ciencia y Tecnolog\'{i}a (FONCICYT) and Direcci\'{o}n General de Asuntos del Personal Academico (DGAPA), Mexico;
Nederlandse Organisatie voor Wetenschappelijk Onderzoek (NWO), Netherlands;
The Research Council of Norway, Norway;
Pontificia Universidad Cat\'{o}lica del Per\'{u}, Peru;
Ministry of Science and Higher Education, National Science Centre and WUT ID-UB, Poland;
Korea Institute of Science and Technology Information and National Research Foundation of Korea (NRF), Republic of Korea;
Ministry of Education and Scientific Research, Institute of Atomic Physics, Ministry of Research and Innovation and Institute of Atomic Physics and Universitatea Nationala de Stiinta si Tehnologie Politehnica Bucuresti, Romania;
Ministerstvo skolstva, vyskumu, vyvoja a mladeze SR, Slovakia;
National Research Foundation of South Africa, South Africa;
Swedish Research Council (VR) and Knut \& Alice Wallenberg Foundation (KAW), Sweden;
European Organization for Nuclear Research, Switzerland;
Suranaree University of Technology (SUT), National Science and Technology Development Agency (NSTDA) and National Science, Research and Innovation Fund (NSRF via PMU-B B05F650021), Thailand;
Turkish Energy, Nuclear and Mineral Research Agency (TENMAK), Turkey;
National Academy of  Sciences of Ukraine, Ukraine;
Science and Technology Facilities Council (STFC), United Kingdom;
National Science Foundation of the United States of America (NSF) and United States Department of Energy, Office of Nuclear Physics (DOE NP), United States of America.
In addition, individual groups or members have received support from:
Czech Science Foundation (grant no. 23-07499S), Czech Republic;
FORTE project, reg.\ no.\ CZ.02.01.01/00/22\_008/0004632, Czech Republic, co-funded by the European Union, Czech Republic;
European Research Council (grant no. 950692), European Union;
Deutsche Forschungs Gemeinschaft (DFG, German Research Foundation) ``Neutrinos and Dark Matter in Astro- and Particle Physics'' (grant no. SFB 1258), Germany;
CONVECS project, CUP C97H23001700002 FESR 2021-2027 program, Italy;
FAIR - Future Artificial Intelligence Research, funded by the NextGenerationEU program (Italy).

\end{acknowledgement}

\bibliographystyle{utphys}   
\bibliography{bibliography}

\newpage
\appendix

%
%

\section{The ALICE Collaboration}
\label{app:collab}
\begin{flushleft} 
\small

I.J.~Abualrob\,\orcidlink{0009-0005-3519-5631}\,$^{\rm 113}$, 
S.~Acharya\,\orcidlink{0000-0002-9213-5329}\,$^{\rm 49}$, 
G.~Aglieri Rinella\,\orcidlink{0000-0002-9611-3696}\,$^{\rm 32}$, 
L.~Aglietta\,\orcidlink{0009-0003-0763-6802}\,$^{\rm 24}$, 
N.~Agrawal\,\orcidlink{0000-0003-0348-9836}\,$^{\rm 25}$, 
Z.~Ahammed\,\orcidlink{0000-0001-5241-7412}\,$^{\rm 133}$, 
S.~Ahmad\,\orcidlink{0000-0003-0497-5705}\,$^{\rm 15}$, 
I.~Ahuja\,\orcidlink{0000-0002-4417-1392}\,$^{\rm 36}$, 
Z.~Akbar$^{\rm 80}$, 
A.~Akindinov\,\orcidlink{0000-0002-7388-3022}\,$^{\rm 139}$
% ...

\section*{Affiliation Notes}

$^{\rm I}$ Deceased\\
$^{\rm II}$ Also at: Max-Planck-Institut fur Physik, Munich, Germany\\
$^{\rm III}$ Also at: Czech Technical University in Prague (CZ)\\
$^{\rm IV}$ Also at: Instituto de Fisica da Universidade de Sao Paulo\\
$^{\rm V}$ Also at: Dipartimento DET del Politecnico di Torino, Turin, Italy\\
$^{\rm VI}$ Also at: Department of Applied Physics, Aligarh Muslim University, Aligarh, India\\
$^{\rm VII}$ Also at: Institute of Theoretical Physics, University of Wroclaw, Poland\\
$^{\rm VIII}$ Also at: Facultad de Ciencias, Universidad Nacional Aut\'{o}noma de M\'{e}xico, Mexico City, Mexico\\
$^{\rm IX}$ Also at: University College of Dublin, Dublin, Ireland\\

\section*{Collaboration Institutes}

$^{1}$ A.I. Alikhanyan National Science Laboratory (Yerevan Physics Institute) Foundation, Yerevan, Armenia\\
$^{2}$ AGH University of Krakow, Cracow, Poland\\
$^{3}$ Bogolyubov Institute for Theoretical Physics, National Academy of Sciences of Ukraine, Kyiv, Ukraine\\
$^{4}$ Bose Institute, Department of Physics  and Centre for Astroparticle Physics and Space Science (CAPSS), Kolkata, India\\
$^{5}$ California Polytechnic State University, San Luis Obispo, California, United States\\
$^{6}$ Central China Normal University, Wuhan, China\\
$^{7}$ Centro de Aplicaciones Tecnol\'{o}gicas y Desarrollo Nuclear (CEADEN), Havana, Cuba\\
$^{8}$ Centro de Investigaci\'{o}n y de Estudios Avanzados (CINVESTAV), Mexico City and M\'{e}rida, Mexico\\
$^{9}$ Chicago State University, Chicago, Illinois, United States\\
$^{10}$ China Nuclear Data Center, China Institute of Atomic Energy, Beijing, China\\
$^{11}$ China University of Geosciences, Wuhan, China\\
$^{12}$ Chungbuk National University, Cheongju, Republic of Korea\\
$^{13}$ Comenius University Bratislava, Faculty of Mathematics, Physics and Informatics, Bratislava, Slovak Republic\\
$^{14}$ Creighton University, Omaha, Nebraska, United States\\
$^{15}$ Department of Physics, Aligarh Muslim University, Aligarh, India\\
$^{16}$ Department of Physics, Pusan National University, Pusan, Republic of Korea\\
$^{17}$ Department of Physics, Sejong University, Seoul, Republic of Korea\\
$^{18}$ Department of Physics, University of California, Berkeley, California, United States\\
$^{19}$ Department of Physics, University of Oslo, Oslo, Norway\\
$^{20}$ Department of Physics and Technology, University of Bergen, Bergen, Norway\\
$^{21}$ Dipartimento di Fisica, Universit\`{a} di Pavia, Pavia, Italy\\
$^{22}$ Dipartimento di Fisica dell'Universit\`{a} and Sezione INFN, Cagliari, Italy\\
$^{23}$ Dipartimento di Fisica dell'Universit\`{a} and Sezione INFN, Trieste, Italy\\
$^{24}$ Dipartimento di Fisica dell'Universit\`{a} and Sezione INFN, Turin, Italy\\
$^{25}$ Dipartimento di Fisica e Astronomia dell'Universit\`{a} and Sezione INFN, Bologna, Italy\\
$^{26}$ Dipartimento di Fisica e Astronomia dell'Universit\`{a} and Sezione INFN, Catania, Italy\\
$^{27}$ Dipartimento di Fisica e Astronomia dell'Universit\`{a} and Sezione INFN, Padova, Italy\\
$^{28}$ Dipartimento di Fisica `E.R.~Caianiello' dell'Universit\`{a} and Gruppo Collegato INFN, Salerno, Italy\\
$^{29}$ Dipartimento DISAT del Politecnico and Sezione INFN, Turin, Italy\\
$^{30}$ Dipartimento di Scienze MIFT, Universit\`{a} di Messina, Messina, Italy\\
$^{31}$ Dipartimento Interateneo di Fisica `M.~Merlin' and Sezione INFN, Bari, Italy\\
$^{32}$ European Organization for Nuclear Research (CERN), Geneva, Switzerland\\
$^{33}$ Faculty of Electrical Engineering, Mechanical Engineering and Naval Architecture, University of Split, Split, Croatia\\
$^{34}$ Faculty of Nuclear Sciences and Physical Engineering, Czech Technical University in Prague, Prague, Czech Republic\\
$^{35}$ Faculty of Physics, Sofia University, Sofia, Bulgaria\\
$^{36}$ Faculty of Science, P.J.~\v{S}af\'{a}rik University, Ko\v{s}ice, Slovak Republic\\
$^{37}$ Faculty of Technology, Environmental and Social Sciences, Bergen, Norway\\
$^{38}$ Frankfurt Institute for Advanced Studies, Johann Wolfgang Goethe-Universit\"{a}t Frankfurt, Frankfurt, Germany\\
$^{39}$ Fudan University, Shanghai, China\\
$^{40}$ Gauhati University, Department of Physics, Guwahati, India\\
$^{41}$ Helmholtz-Institut f\"{u}r Strahlen- und Kernphysik, Rheinische Friedrich-Wilhelms-Universit\"{a}t Bonn, Bonn, Germany\\
$^{42}$ Helsinki Institute of Physics (HIP), Helsinki, Finland\\
$^{43}$ High Energy Physics Group,  Universidad Aut\'{o}noma de Puebla, Puebla, Mexico\\
$^{44}$ Horia Hulubei National Institute of Physics and Nuclear Engineering, Bucharest, Romania\\
$^{45}$ HUN-REN Wigner Research Centre for Physics, Budapest, Hungary\\
$^{46}$ Indian Institute of Technology Bombay (IIT), Mumbai, India\\
$^{47}$ Indian Institute of Technology Indore, Indore, India\\
$^{48}$ INFN, Laboratori Nazionali di Frascati, Frascati, Italy\\
$^{49}$ INFN, Sezione di Bari, Bari, Italy\\
$^{50}$ INFN, Sezione di Bologna, Bologna, Italy\\
$^{51}$ INFN, Sezione di Cagliari, Cagliari, Italy\\
$^{52}$ INFN, Sezione di Catania, Catania, Italy\\
$^{53}$ INFN, Sezione di Padova, Padova, Italy\\
$^{54}$ INFN, Sezione di Pavia, Pavia, Italy\\
$^{55}$ INFN, Sezione di Torino, Turin, Italy\\
$^{56}$ INFN, Sezione di Trieste, Trieste, Italy\\
$^{57}$ Inha University, Incheon, Republic of Korea\\
$^{58}$ Institute for Gravitational and Subatomic Physics (GRASP), Utrecht University/Nikhef, Utrecht, Netherlands\\
$^{59}$ Institute of Experimental Physics, Slovak Academy of Sciences, Ko\v{s}ice, Slovak Republic\\
$^{60}$ Institute of Physics, Homi Bhabha National Institute, Bhubaneswar, India\\
$^{61}$ Institute of Physics of the Czech Academy of Sciences, Prague, Czech Republic\\
$^{62}$ Institute of Space Science (ISS), Bucharest, Romania\\
$^{63}$ Institut f\"{u}r Kernphysik, Johann Wolfgang Goethe-Universit\"{a}t Frankfurt, Frankfurt, Germany\\
$^{64}$ Instituto de Ciencias Nucleares, Universidad Nacional Aut\'{o}noma de M\'{e}xico, Mexico City, Mexico\\
$^{65}$ Instituto de F\'{i}sica, Universidade Federal do Rio Grande do Sul (UFRGS), Porto Alegre, Brazil\\
$^{66}$ Instituto de F\'{\i}sica, Universidad Nacional Aut\'{o}noma de M\'{e}xico, Mexico City, Mexico\\
$^{67}$ iThemba LABS, National Research Foundation, Somerset West, South Africa\\
$^{68}$ Jeonbuk National University, Jeonju, Republic of Korea\\
$^{69}$ Korea Institute of Science and Technology Information, Daejeon, Republic of Korea\\
$^{70}$ Laboratoire de Physique Subatomique et de Cosmologie, Universit\'{e} Grenoble-Alpes, CNRS-IN2P3, Grenoble, France\\
$^{71}$ Lawrence Berkeley National Laboratory, Berkeley, California, United States\\
$^{72}$ Lund University Department of Physics, Division of Particle Physics, Lund, Sweden\\
$^{73}$ Marietta Blau Institute, Vienna, Austria\\
$^{74}$ Nagasaki Institute of Applied Science, Nagasaki, Japan\\
$^{75}$ Nara Women{'}s University (NWU), Nara, Japan\\
$^{76}$ National and Kapodistrian University of Athens, School of Science, Department of Physics , Athens, Greece\\
$^{77}$ National Centre for Nuclear Research, Warsaw, Poland\\
$^{78}$ National Institute of Science Education and Research, Homi Bhabha National Institute, Jatni, India\\
$^{79}$ National Nuclear Research Center, Baku, Azerbaijan\\
$^{80}$ National Research and Innovation Agency - BRIN, Jakarta, Indonesia\\
$^{81}$ Niels Bohr Institute, University of Copenhagen, Copenhagen, Denmark\\
$^{82}$ Nikhef, National institute for subatomic physics, Amsterdam, Netherlands\\
$^{83}$ Nuclear Physics Group, STFC Daresbury Laboratory, Daresbury, United Kingdom\\
$^{84}$ Nuclear Physics Institute of the Czech Academy of Sciences, Husinec-\v{R}e\v{z}, Czech Republic\\
$^{85}$ Oak Ridge National Laboratory, Oak Ridge, Tennessee, United States\\
$^{86}$ Ohio State University, Columbus, Ohio, United States\\
$^{87}$ Physics department, Faculty of science, University of Zagreb, Zagreb, Croatia\\
$^{88}$ Physics Department, Panjab University, Chandigarh, India\\
$^{89}$ Physics Department, University of Jammu, Jammu, India\\
$^{90}$ Physics Program and International Institute for Sustainability with Knotted Chiral Meta Matter (WPI-SKCM$^{2}$), Hiroshima University, Hiroshima, Japan\\
$^{91}$ Physikalisches Institut, Eberhard-Karls-Universit\"{a}t T\"{u}bingen, T\"{u}bingen, Germany\\
$^{92}$ Physikalisches Institut, Ruprecht-Karls-Universit\"{a}t Heidelberg, Heidelberg, Germany\\
$^{93}$ Physik Department, Technische Universit\"{a}t M\"{u}nchen, Munich, Germany\\
$^{94}$ Politecnico di Bari and Sezione INFN, Bari, Italy\\
$^{95}$ Research Division and ExtreMe Matter Institute EMMI, GSI Helmholtzzentrum f\"ur Schwerionenforschung GmbH, Darmstadt, Germany\\
$^{96}$ Saga University, Saga, Japan\\
$^{97}$ Saha Institute of Nuclear Physics, Homi Bhabha National Institute, Kolkata, India\\
$^{98}$ School of Physics and Astronomy, University of Birmingham, Birmingham, United Kingdom\\
$^{99}$ Secci\'{o}n F\'{\i}sica, Departamento de Ciencias, Pontificia Universidad Cat\'{o}lica del Per\'{u}, Lima, Peru\\
$^{100}$ SUBATECH, IMT Atlantique, Nantes Universit\'{e}, CNRS-IN2P3, Nantes, France\\
$^{101}$ Sungkyunkwan University, Suwon City, Republic of Korea\\
$^{102}$ Suranaree University of Technology, Nakhon Ratchasima, Thailand\\
$^{103}$ Technical University of Ko\v{s}ice, Ko\v{s}ice, Slovak Republic\\
$^{104}$ The Henryk Niewodniczanski Institute of Nuclear Physics, Polish Academy of Sciences, Cracow, Poland\\
$^{105}$ The University of Texas at Austin, Austin, Texas, United States\\
$^{106}$ Universidad Aut\'{o}noma de Sinaloa, Culiac\'{a}n, Mexico\\
$^{107}$ Universidade de S\~{a}o Paulo (USP), S\~{a}o Paulo, Brazil\\
$^{108}$ Universidade Estadual de Campinas (UNICAMP), Campinas, Brazil\\
$^{109}$ Universidade Federal do ABC, Santo Andre, Brazil\\
$^{110}$ Universitatea Nationala de Stiinta si Tehnologie Politehnica Bucuresti, Bucharest, Romania\\
$^{111}$ University of Cape Town, Cape Town, South Africa\\
$^{112}$ University of Derby, Derby, United Kingdom\\
$^{113}$ University of Houston, Houston, Texas, United States\\
$^{114}$ University of Jyv\"{a}skyl\"{a}, Jyv\"{a}skyl\"{a}, Finland\\
$^{115}$ University of Kansas, Lawrence, Kansas, United States\\
$^{116}$ University of Liverpool, Liverpool, United Kingdom\\
$^{117}$ University of Science and Technology of China, Hefei, China\\
$^{118}$ University of Silesia in Katowice, Katowice, Poland\\
$^{119}$ University of South-Eastern Norway, Kongsberg, Norway\\
$^{120}$ University of Tennessee, Knoxville, Tennessee, United States\\
$^{121}$ University of the Witwatersrand, Johannesburg, South Africa\\
$^{122}$ University of Tokyo, Tokyo, Japan\\
$^{123}$ University of Tsukuba, Tsukuba, Japan\\
$^{124}$ Universit\"{a}t M\"{u}nster, Institut f\"{u}r Kernphysik, M\"{u}nster, Germany\\
$^{125}$ Universit\'{e} Clermont Auvergne, CNRS/IN2P3, LPC, Clermont-Ferrand, France\\
$^{126}$ Universit\'{e} de Lyon, CNRS/IN2P3, Institut de Physique des 2 Infinis de Lyon, Lyon, France\\
$^{127}$ Universit\'{e} de Strasbourg, CNRS, IPHC UMR 7178, F-67000 Strasbourg, France, Strasbourg, France\\
$^{128}$ Universit\'{e} Paris-Saclay, Centre d'Etudes de Saclay (CEA), IRFU, D\'{e}partment de Physique Nucl\'{e}aire (DPhN), Saclay, France\\
$^{129}$ Universit\'{e}  Paris-Saclay, CNRS/IN2P3, IJCLab, Orsay, France\\
$^{130}$ Universit\`{a} degli Studi di Foggia, Foggia, Italy\\
$^{131}$ Universit\`{a} del Piemonte Orientale, Vercelli, Italy\\
$^{132}$ Universit\`{a} di Brescia, Brescia, Italy\\
$^{133}$ Variable Energy Cyclotron Centre, Homi Bhabha National Institute, Kolkata, India\\
$^{134}$ Warsaw University of Technology, Warsaw, Poland\\
$^{135}$ Wayne State University, Detroit, Michigan, United States\\
$^{136}$ Yale University, New Haven, Connecticut, United States\\
$^{137}$ Yildiz Technical University, Istanbul, Turkey\\
$^{138}$ Yonsei University, Seoul, Republic of Korea\\
$^{139}$ Affiliated with an institute formerly covered by a cooperation agreement with CERN\\
$^{140}$ Affiliated with an international laboratory covered by a cooperation agreement with CERN.\\

\end{flushleft} 

\end{document}